\documentclass[journal]{IEEEtran}

\usepackage[backend=bibtex8,sorting=none]{biblatex}

\bibliography{newlib}

\title{On the Relationship Between Scintillation Anisotropy and Crystal Structure in Pure Crystalline Organic Scintillator Materials} 
\author{Patricia~Schuster,~\IEEEmembership{Member,~IEEE,} Patrick~Feng,~\IEEEmembership{Member,~IEEE,} and~Erik~Brubaker,~\IEEEmembership{Member,~IEEE,}%
\thanks{P.~Schuster is with the Nuclear Engineering and Radiological Sciences Department, University of Michigan, Ann Arbor, MI, 48105 USA e-mail: pfschus@umich.edu.}%
\thanks{P.~Feng and E.~Brubaker are with Sandia National Laboratories, Livermore, CA, 94550 USA.}%
\thanks{Manuscript received November 21, 2017.}
}

\usepackage{url}                                        
\usepackage{xcolor,colortbl}                            
\usepackage{amssymb}                                    
\usepackage{amsmath,amsthm}
\usepackage{amsthm}                                     
\usepackage{epstopdf}                                   
\usepackage[caption=false,font=footnotesize]{subfig}    
\usepackage{graphicx}
\usepackage{float}
\usepackage{placeins}
\usepackage{multirow}                                   
\usepackage{etoolbox}                                   
\usepackage[hidelinks]{hyperref}                        
\hypersetup{                    
	colorlinks,
	citecolor=black,
	filecolor=black,
	linkcolor=black,
	urlcolor=black
}
\usepackage{verbatim}                                   
\usepackage[makeroom]{cancel}                           
\usepackage[utf8]{inputenc}                             
\usepackage{listings}                                   
\usepackage[framed]{matlab-prettifier}                  
\usepackage{enumitem}									
\usepackage{easy}             							
\usepackage{nomencl}									
\makenomenclature 
\usepackage{xspace}

\newcommand{\fig}[1]     {Fig.~\ref{#1}}
\newcommand{\tab}[1]     {Table~\ref{#1}}

\newcommand{\secref}[1]  {Sec.~\ref{#1}}
\newcommand{\eqnref}[1]  {Eq.~\eqref{#1}}

\newcommand{\etal}       {\textit{et~al.}\xspace}

\DeclareMathOperator\erf{erf}

\newcommand{\plotsdescribed}{Plots described at beginning of \secref{sec:results}.\xspace}

\newcommand{\Sgrd}       {\ensuremath{S_0}\xspace}
\newcommand{\Sone}       {\ensuremath{S_1}\xspace}
\newcommand{\Tone}       {\ensuremath{T_1}\xspace}

\newcommand{\rhoS}       {\ensuremath{\rho_S}\xspace}
\newcommand{\rhoT}       {\ensuremath{\rho_T}\xspace}

\newcommand{\ALO}        {\ensuremath{A_{\LOave}}\xspace}
\newcommand{\ATTT}       {\ensuremath{A_{\TTTave}}\xspace}
\newcommand{\AP}         {\ensuremath{A_{\Pave}}\xspace}
\newcommand{\AD}         {\ensuremath{A_{\Dave}}\xspace}

\newcommand{\shd}{\cellcolor{lightgray}}

\newcommand{\LO}         {\ensuremath{L}\xspace}

\newcommand{\TTT}        {\ensuremath{S}\xspace}
\newcommand{\LOave}      {\ensuremath{\hat{\LO}}\xspace}
\newcommand{\TTTave}     {\ensuremath{\hat{\TTT}}\xspace}
\newcommand{\Pave}     {\ensuremath{\hat{P}}\xspace}
\newcommand{\Dave}     {\ensuremath{\hat{D}}\xspace}
\newcommand{\LOaverange} {\ensuremath{\hat{\LO}\pm\sigma}\xspace}

\newcommand{\Pmax}     {\ensuremath{\hat{P}_{\textrm{max}}}\xspace}
\newcommand{\Dmax}     {\ensuremath{\hat{D}_{\textrm{max}}}\xspace}
\newcommand{\Pmin}     {\ensuremath{\hat{P}_{\textrm{min}}}\xspace}
\newcommand{\Dmin}     {\ensuremath{\hat{D}_{\textrm{min}}}\xspace}

\newcommand{\dEdx}       {\ensuremath{\mathrm{d}E/\mathrm{d}x}\xspace}
\newcommand{\En}    {\ensuremath{E_{\textrm{n}}}\xspace}

\newcommand{\genunit}[2]{\ensuremath{#1~\text{#2}}\xspace}

\newcommand{\keVee}[1]  {\genunit{#1}{keVee}}
\newcommand{\MeV}[1]    {\genunit{#1}{MeV}}

\newcommand{\degrees}[1]{\ensuremath{#1^{\mathrm{o}}}\xspace}
\newcommand{\degreesC}[1]{\ensuremath{#1^{\mathrm{o}}\textrm{C}}\xspace}

\newcommand{\Csots}{\ensuremath{^{137}\text{Cs}}\xspace}
\newcommand{\Natt} {\ensuremath{^{22}\text{Na}}\xspace}

\begin{document}
\maketitle

\begin{abstract}
	The scintillation anisotropy effect for proton recoil events has been investigated in five pure organic crystalline materials: anthracene, \textit{trans}-stilbene, p-terphenyl, bibenzyl, and diphenylacetylene. These measurements include characterization of the scintillation response for one hemisphere of proton recoil directions in each crystal. In addition to standard measurements of the total light output and pulse shape at each angle, the prompt and delayed light anisotropies are analyzed, allowing for investigation of the singlet and triplet molecular excitation behaviors independently. This work provides new quantitative and qualitative observations that make progress toward understanding the physical mechanisms behind the scintillation anisotropy. These measurements show that the relationship between the prompt and delayed light anisotropies is correlated with crystal structure, as it changes between the pi-stacked crystal structure materials (anthracene and p-terphenyl) and the herringbone crystal structure materials (stilbene, bibenzyl, and diphenylacetylene). The observations are consistent with a model in which there are preferred directions of kinetic processes for the molecular excitations. These processes and the impact of their directional dependencies on the scintillation anisotropy are discussed.
\end{abstract}

\IEEEpeerreviewmaketitle

\printnomenclature[.5in]

\section{Introduction}

\IEEEPARstart{O}{rganic} single crystals are promising materials as ionizing radiation detectors, especially when characterizing mixed radiation fields. Their rich hydrogen content makes it possible to detect both fast neutrons and gamma-rays, between which they can discriminate using pulse shape discrimination (PSD). While these materials have been available for decades, recent advancements in methods for growing organic single crystals from solution processing have produced large-scale crystalline materials (on the order of 10 cm in each dimension) with superior scintillation properties~\cite{Fraboni2016,Zaitseva2015}, igniting renewed interest in their use. 

PSD is possible due to the fact that the light emission generated in response to interactions from different particle types varies in its time distribution~\cite{Wright1953,Brooks1957b}. The energy resolution and PSD performance varies across organic scintillator materials~\cite{Yanagida2015,Carlson2016}, largely due to differences in their structure and material properties. Recently, the \textit{trans} isomer of stilbene has become a favorite as its production via solution processing produces crystals with superior PSD performance~\cite{Fraboni2016}. Stilbene is also preferred in many applications as an alternative to organic liquid scintillators, which have practical challenges including leaking, expansion with temperature, and oxidation.

In many crystalline organic scintillator materials, the light output response varies with heavy charged particle interaction direction. This scintillation anisotropy effect was first observed with alpha particles~\cite{Heckmann1959,Heckmann1961a,Heckmann1961,W.F.Kienzle1961} and later with proton~\cite{Tsukada1962,Tsukada1965,Brooks1974} and carbon recoils~\cite{Oliver1968,Sekiya2003} generated by neutron interactions. In a neutron detection event, the total amount of light emitted and the time distribution of the light emission have been observed to vary with the direction of the proton recoil following the neutron-proton scatter interaction. The directional effect under discussion here is experimentally independent of and distinct from any anisotropy in the direction of scintillation light emission. For many applications in which the proton recoil direction is unknown, this effect degrades the light output resolution and PSD performance.

The following is a summary of the current state of understanding surrounding the scintillation anisotropy effect; a more detailed discussion can be found in~\cite{SchusterThesis2016}:

\begin{itemize}
	\item The magnitude of the anisotropy varies with heavy charged particle energy~\cite{Heckmann1959,Heckmann1961,Tsukada1965,Brooks1974,Schuster2016}.
	\item The scintillation anisotropy has been shown to vary in magnitude and behavior across materials, where behavior includes the relationship between the light output and pulse shape anisotropies~\cite{Heckmann1961a, W.F.Kienzle1961,Brooks1974,Brubaker2010}.
	\item The effect is not due to channeling or anisotropy in \dEdx~\cite{Tsukada1965,Brooks1974}.
	\item No anisotropy is present for electron recoil events generated by gamma-ray interactions, as noted in~\cite{Heckmann1959,Brooks1974}. A previous paper by the authors demonstrated experimentally that no anisotropy is observed for electron recoils and muon interactions, supporting the hypothesis that a high \dEdx is required to produce the effect~\cite{Schuster2016}. 
	\item Further work by the authors presented measurements of the effect in stilbene and demonstrated that the scintillation anisotropy is approximately uniform across multiple crystals with different shapes and growth methods, indicating that the effect does not depend heavily on crystal quality or bulk crystal geometry~\cite{Schuster2016a}.
	\item The authors also confirmed by measurements that no anisotropy is observed in amorphous liquid or plastic materials~\cite{Schuster2016a}, demonstrating that the effect is in fact due to crystal structure, and not due to variations in light collection efficiency or an external effect on the detector system.
\end{itemize}
\nomenclature{\dEdx}{Stopping power, or specific energy loss}
 
For users who wish to use organic crystal scintillators as radiation detectors, the scintillation anisotropy may add a new challenge, as it introduces significant variation in the light output and pulse shapes of events at different crystal axes directions. Alternately, the scintillation anisotropy may present an opportunity to employ these materials as directional detectors, as investigated by \citeauthor{Brubaker2010}~\cite{Brubaker2010}, or to detect WIMP dark matter~\cite{Sekiya2003,Belli1992}. Additionally, the scintillation anisotropy is a function of many poorly characterized physical processes in these materials and further investigating the effect may provide greater insight on those processes.

This paper contributes to the body of knowledge surrounding the anisotropy in three major ways:

First, by measuring the effect in five single crystal materials with the same measurement system, allowing for consistent characterization of the effect. While the effect has been measured in several materials by numerous authors~\cite{Heckmann1961a,W.F.Kienzle1961,Tsukada1962,Tsukada1965,Brooks1974,Brubaker2010}, inconsistencies in the measurement systems and analysis techniques limit comparisons between each author's work.
 
Second, the anisotropy characterization includes measurements at more than 70 proton recoil directions across a hemisphere in each material. This extends previous authors' work that was often limited to measurements only at the presumed maximum and minimum directions. 

Third, by measuring the anisotropies of the prompt and delayed light components separately. These components roughly correspond to the behavior of the initial populations of singlet and triplet excitations, respectively. In most previously published measurements of this effect, the scintillation anisotropy is measured with respect to the total amount of light emitted, often referred to as the light output, and the time distribution of light emitted, often expressed as a one-dimensional pulse shape parameter. While these are both useful quantities, they do not isolate the behavior of the initial singlet and triplet excitation populations.

These measurements will be analyzed quantitatively, qualitatively, and with respect to the crystal structures in order to determine whether there is a relationship between a crystalline material's scintillation anisotropy and its crystal structure.

\section{Materials and Experimental Setup}

\subsection{Crystal material preparation}

This paper reports measurements on five pure crystalline organic scintillator materials: anthracene, \textit{trans}-stilbene, p-terphenyl, bibenzyl, and diphenylacetylene (DPAC). The anthracene sample was the same one measured in~\cite{Schuster2016} and is an older sample with considerable wear. The p-terphenyl sample was of approximately the same quality and also has unknown history. Both of these samples have been polished many times and were produced by the melt-growth technique. The stilbene (Cubic B from~\cite{Schuster2016a}), bibenzyl, and DPAC samples were provided by Lawrence Livermore National Laboratory and were grown with a solution-growth method. These samples were all very high quality with few imperfections on the surface. The comparison of the anisotropy effect across all samples will not attempt to account for the differences in quality among these samples as it was demonstrated in previous work that stilbene samples of different sizes and growth methods produced approximately the same scintillation anisotropy in magnitude and behavior~\cite{Schuster2016a}. Characteristics of the five crystals reported in this paper are provided in~\tab{tab:materials}.

\begin{table*}[]
	\footnotesize
	\centering
	\caption{Summary of physical characteristics of organic crystal scintillator materials measured in this paper.}
	\label{tab:materials}
	\begin{tabular}{lllllll}
		Material                & Formula          & Density (g/cm$^3$) & Growth   & Shape             & Dimensions (cm)        & Volume (cm$^3$)           \\
		\hline
		Anthracene              & C$_{14}$H$_{10}$ & 1.28               & Melt     & Cylinder          & h = 1.9, d = 3.2       & 15.1    \\
		Stilbene          & C$_{14}$H$_{12}$ & 1.16               & Solution & Rectangular Prism & 1.9 x 1.9 x 1.9        & 6.9     \\
		P-terphenyl             & C$_{18}$H$_{14}$ & 1.24               & Melt     & Cylinder          & h = 2.5, d = 2.6       & 13.3    \\
		Bibenzyl                & C$_{14}$H$_{10}$ & 0.99               & Solution & Rectangular Prism & 0.6 x 2.3 x 3.2        & 4.5     \\
		Diphenylacetylene       & C$_{14}$H$_{14}$ & 0.98               & Solution & Trapezoidal Prism & 1.0 x 2.7 x (2.3, 3.2) & 7.4      
	\end{tabular}
\end{table*}

\begin{figure}
	\centering
	\includegraphics[trim={0cm 0cm 0cm 0cm},clip,width=.7\columnwidth]{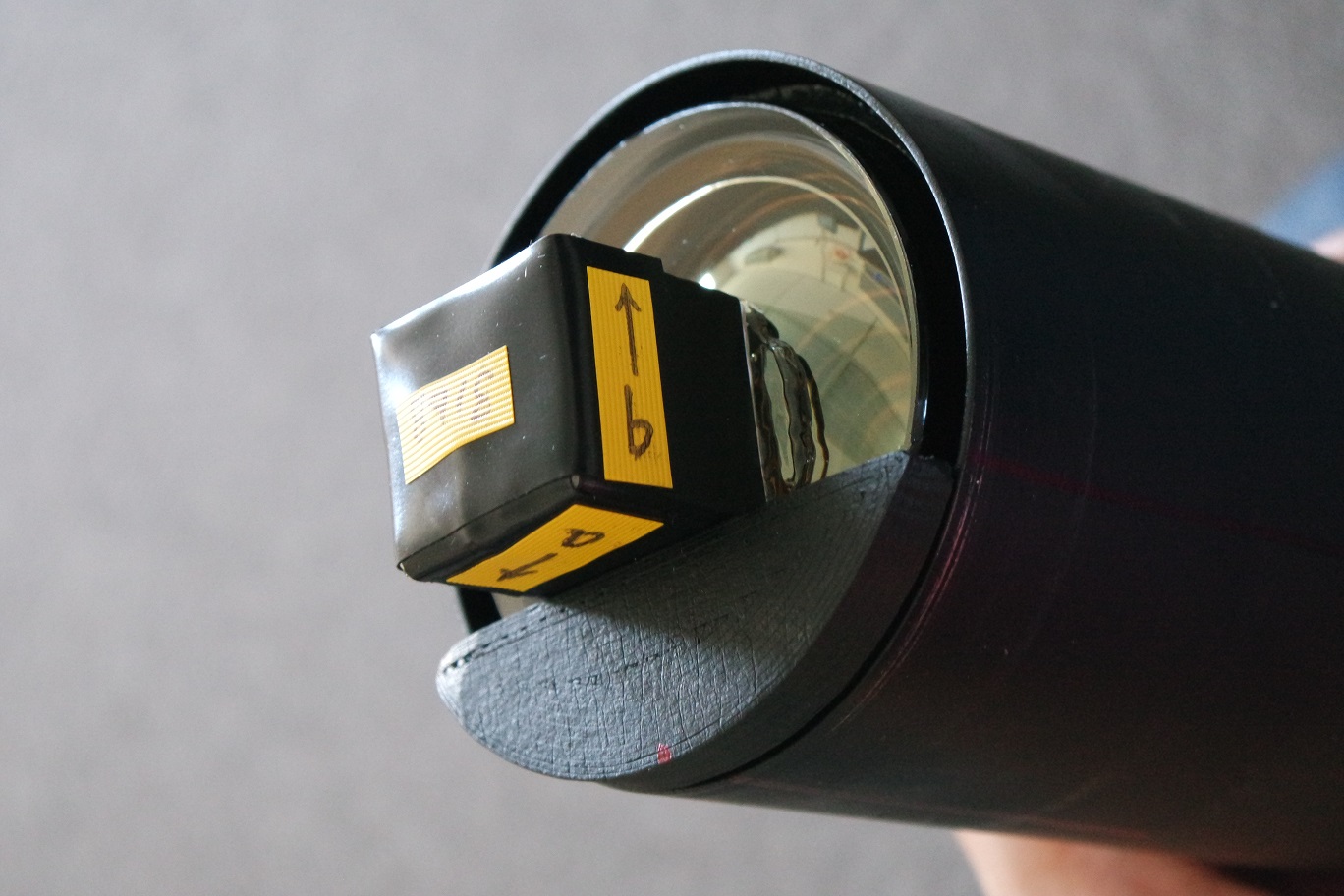}
	\caption{Wrapped stilbene crystal positioned against a plastic guide and mounted to the PMT face with optical grease.
		\label{fig:stil_mounted}
	}
\end{figure}

Each crystal was polished and wrapped in several layers of teflon tape to improve light collection. In some cases, a layer of electrical tape was wrapped around the teflon tape for easier handling. The crystals were then mounted on the face of a 60 mm Hamamatsu H1949-50 photomultiplier tube (PMT) assembly using V-788 optical grease. Each crystal was positioned against a straight plastic guide to prevent the crystal from moving during the assembly, as shown for the stilbene crystal in~\fig{fig:stil_mounted}. A black plastic cap was placed over the crystal and PMT and then wrapped with several layers of black tape to block external light. The high voltage was adjusted separately for each detector in order to best use the dynamic range of the digitizer. 

\subsection{Rotational measurement system}\label{sec:rot_system}

For the measurements presented here, a Thermo Scientific MP320 neutron generator was used to produce roughly monoenergetic neutrons with $\En=\MeV{14.1}$ (via the DT reaction) and $\En=\MeV{2.5}$ (via the DD reaction). In order to measure the light emission in response to proton recoil events at known energies and directions in the crystal axes, only full-energy interactions are used. In these interactions, a neutron deposits its full energy onto the proton and the proton recoil travels in the forward direction. In each measurement, the orientation between the detector and neutron generator was fixed so that one proton recoil direction could be characterized. For each crystal, measurements were taken at many angles in order to characterize the response to many proton recoil directions in the crystal axes. 

\begin{figure}
	\centering
	\includegraphics[trim={0cm 0cm 0cm 0cm},clip,width=\columnwidth]{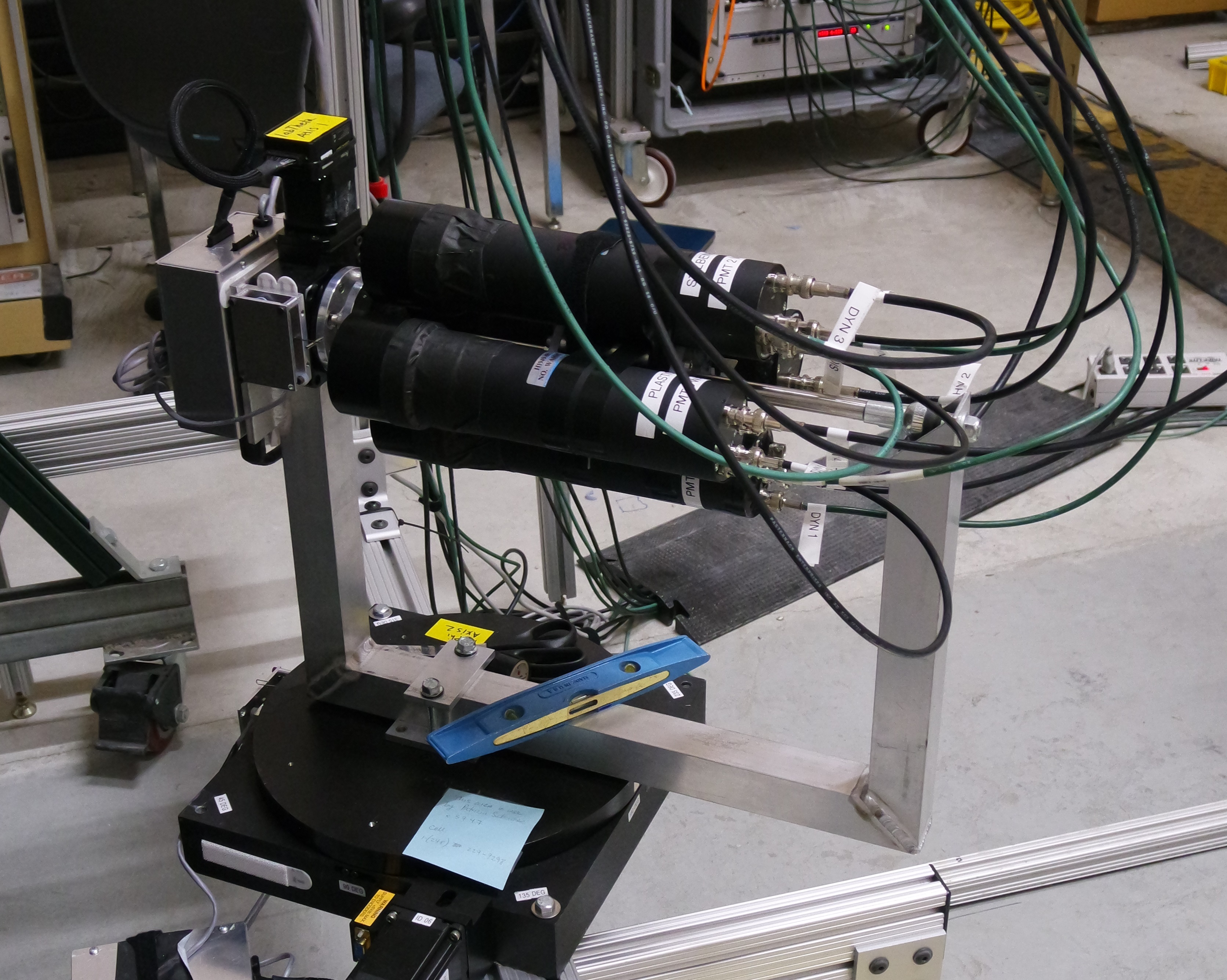}
	\caption{Four detectors assembled on the rotational stage.
		\label{fig:dets_mounted}
	}
\end{figure}

The detectors were mounted to a motor-driven rotational stage, which allowed for measurements of up to four detectors at a time and automated positioning at fixed angles.  Multiple sets of measurements were performed in order to characterize the five pure crystalline materials studied in this work.  \fig{fig:dets_mounted} shows four detectors assembled on the stage. This stage has two axes of rotation and can position each of the four detectors at any angle in $4\pi$. In this work, only $2\pi$ of angles were measured in each crystal due to crystal symmetry.

\begin{figure}[!t]
	\centering
	\subfloat[Three dimensions]{\includegraphics[trim={3.7cm 3cm 2.8cm 3cm},clip,width=2.8in]{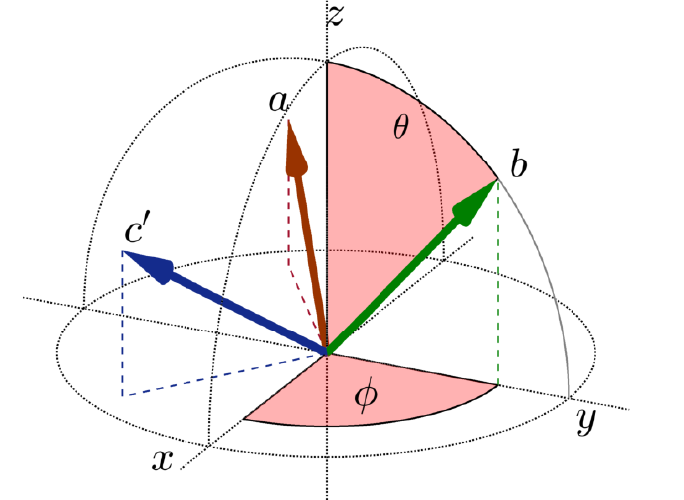}%
		\label{fig:axes_3d}}
	\hfil
	\subfloat[Two dimensions with $r=\sqrt{1-\cos{\theta}}$]{\includegraphics[trim={0cm 0cm 0cm 0cm},clip,width=2in]{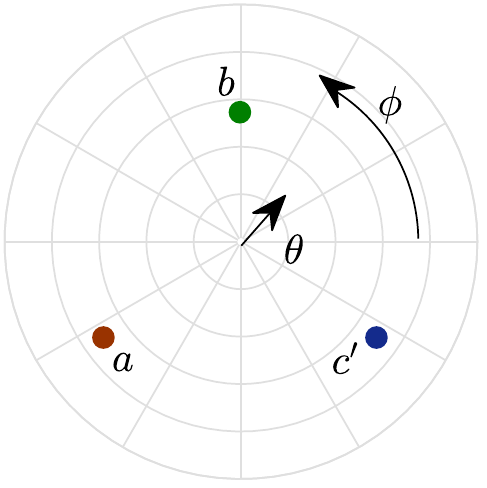}%
		\label{fig:axes_2d}}
	\caption{Visualization of proton recoil directions within a hemisphere of space in (a) three dimensions and (b) two dimensions. Vectors $a$, $b$, and $c'$ correspond to the known axes directions in the measured stilbene sample. The angles $\phi$ and $\theta$ describe the vector directions in spherical coordinates, where $0^\circ < \phi < 360^\circ$ and $0^\circ < \theta < 90^\circ$.}
	\label{fig:axes_defs}
\end{figure}

\nomenclature{($\phi,\theta$)}{Proton recoil direction in spherical coordinates}

The anisotropy effect in the light output, pulse shape, prompt light, and delayed light will each be visualized using a two-dimensional distribution that collapses the hemisphere of measured proton recoil directions onto a flat surface. \fig{fig:axes_defs} shows the technique for visualizing three-dimensional vector directions as points on a two-dimensional distribution. The directions labeled $a$, $b$, and $c'$ correspond to known crystal axis directions in the stilbene sample. The axes for visualization were oriented such that the $a$, $b$, and $c'$ axes, which correspond to qualitative features in the anisotropy, are on the interior of the two-dimensional plot. The stilbene sample is the only crystal in which the crystal axis directions were known. In each other crystal, an arbitrary set of axes was established and oriented such that the features in each expected light output distribution were in approximately the same location as in stilbene for easy comparison across materials.

\section{Analysis Methods}

\subsection{Basic pulse processing}

\begin{figure*}
	\centering
	\includegraphics[trim={0cm 0cm 0cm 0cm},clip,width=.6\textwidth]{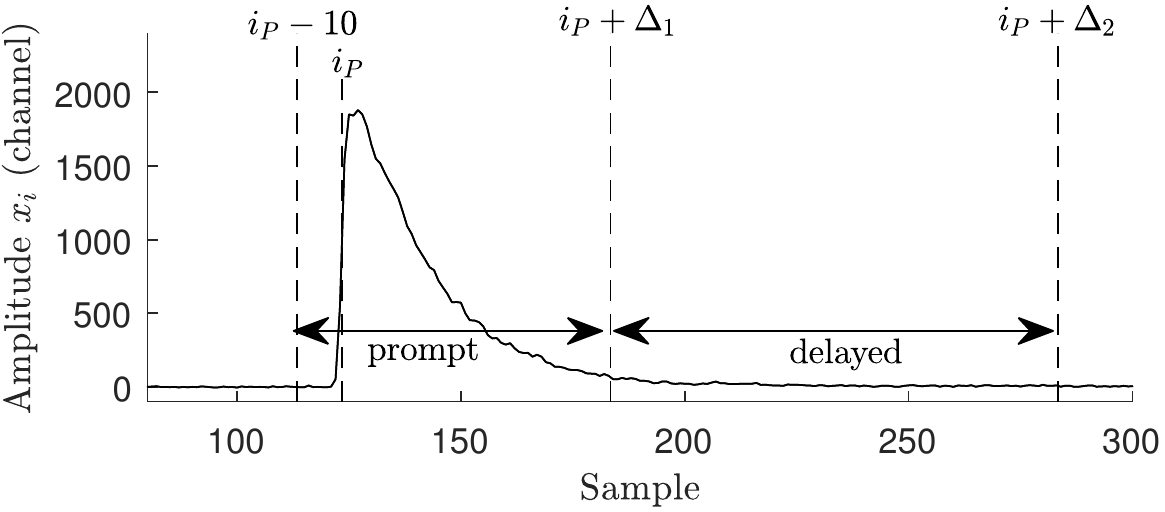}
	\caption{Illustration of a baseline-subtracted raw pulse in anthracene with timing windows for prompt and delayed light components indicated.
		\label{fig:raw_pulse}
	}
\end{figure*}

Pulses are recorded by a Struck SIS3350 500~MHz digitizer with 384 samples per pulse. Raw pulses are recorded and then baseline-subtracted using an average of 85 samples. \fig{fig:raw_pulse} shows a baseline-subtracted pulse, where $x_i$ is the amplitude of each sample measured in digitizer channel units. For each pulse, the total light output \LO is calculated by~\eqnref{eqn:L}. The light output to keV energy equivalent (keVee) conversion is performed by calibrating with a \Csots or \Natt gamma-ray source. This calibration provides a rough scale of light output (in keVee) but is not exact, as the calibration was often performed in a region on the lower end of the digitizer's dynamic range and then applied to events in a higher region of the dynamic range. A pulse shape parameter \TTT is calculated by~\eqnref{eqn:S}. This pulse shape parameter \TTT is often referred to as the ``tail-to-total'' value and provides a measurement of the time distribution of the light emission. The ``tail'' region may be considered the delayed light emission, and the ``total'' region is the sum of what can be considered the prompt and delayed emissions, as labeled in~\fig{fig:raw_pulse}. 
\nomenclature{$x_i$}{Pulse sample amplitude in digitizer channel units}%
\nomenclature{\LO}{Light output of a single event}%
\nomenclature{\TTT}{Pulse shape parameter of a single event}%
\nomenclature{$i_p$}{Sub-sample position at which pulse amplitude exceeds 50\% of its maximum}

\begin{equation}\label{eqn:L}
\LO = \Sigma_{i = 1}^{i = 384} x_i
\end{equation}

\begin{equation}\label{eqn:S}
\TTT = \frac{\Sigma_{i_P+\Delta_1}^{i_P+\Delta_2} x_i}{\Sigma_{i_P-10}^{i_P+\Delta_2} x_i}
\end{equation}

The prompt and delayed time regions are defined by boundaries $\Delta_1$ and $\Delta_2$ steps after $i_p$, the sub-sample position at which the amplitude of the pulse exceeds 50\% of its maximum value. The $\Delta$ values for each material are shown in \tab{tab:deltas}. These $\Delta$ values were optimized for maximal separation between the neutron and gamma-ray regions in a distribution of \LO~vs.~\TTT values measured for the mixed field of radiation produced by a DT neutron generator. 

\nomenclature{$\Delta_1$,$\Delta_2$}{Step sizes in number of samples for defining prompt and delayed regions in a raw pulse}

\begin{table}[]
	\footnotesize
	\centering
	\caption{Step values used in the calculation of prompt and delayed window boundaries for each material. Units are digitizer samples, measured at 2~ns intervals.}
	\label{tab:deltas}
	\begin{tabular}{lll}
		Material              & $\Delta_1$ & $\Delta_2$ \\
		\hline
		Anthracene               & 60      & 160     \\
		Stilbene                 & 10      & 60      \\
		P-Terphenyl              & 6       & 60      \\
		Bibenzyl (BB)            & 24      & 100     \\
		Diphenylacetylene (DPAC) & 6       & 80     
	\end{tabular}
\end{table}

\subsection{Light emission characterization}

\begin{figure*}
	\centering
	\includegraphics[trim={0cm 0cm 0cm 0cm},clip,width=.75\textwidth]{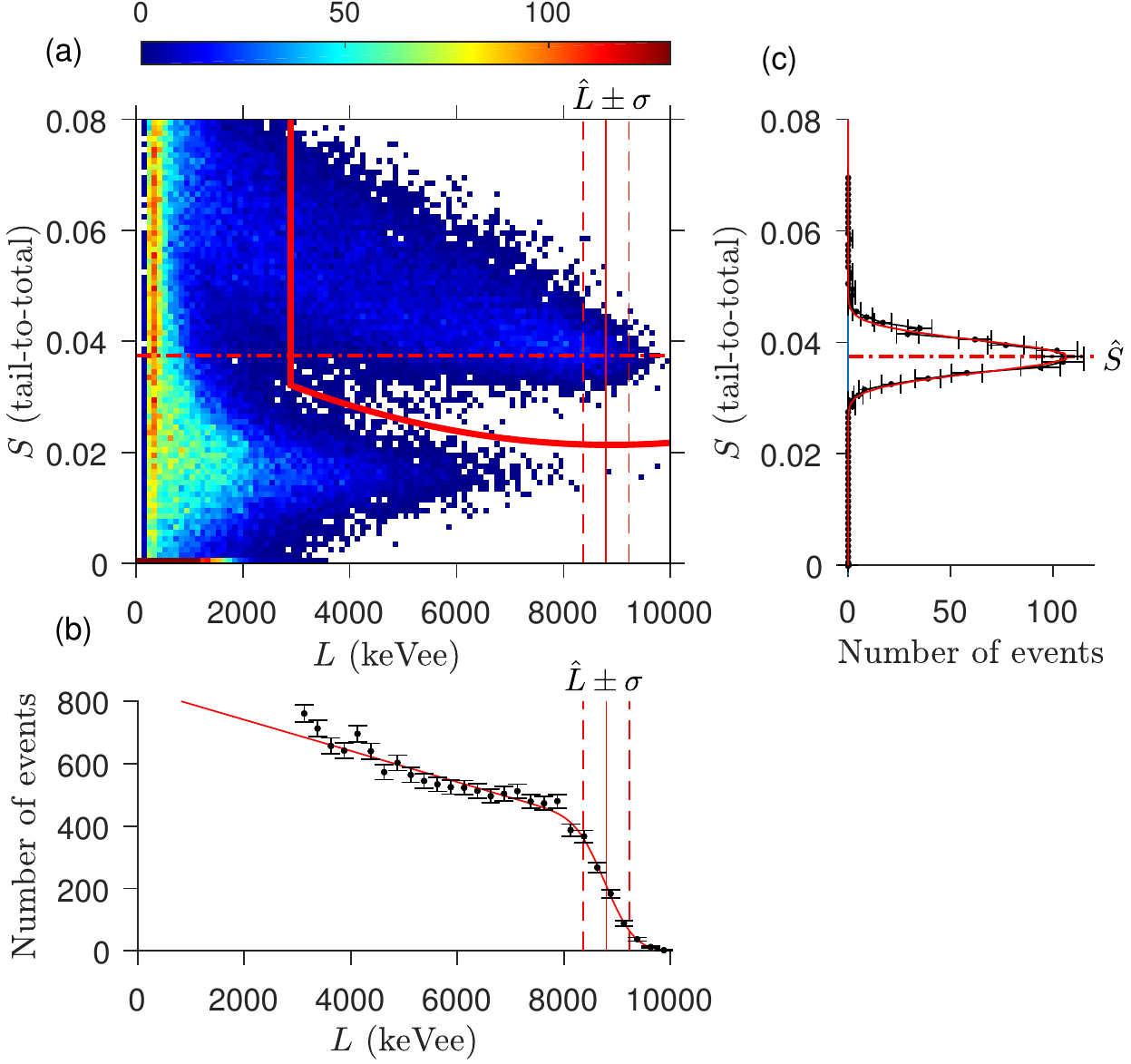}
	\caption{Distributions produced for calculating \LOave and \TTTave for \MeV{14.1} proton recoil events in anthracene. (a) Two-dimensional histogram of \LO~vs.~\TTT values for all events. Events above and to the right of the bold red curve are classified as neutron events. (b) Light output spectrum of neutron events. The vertical solid and dashed lines in (a) and (b) indicate the range $\LOave\pm\sigma$ that includes events  classified as full-energy neutron interactions. (c) Distribution of pulse shape values for full-energy neutron interactions, displayed with inverted axes to correspond with (a). The horizontal dot-dashed lines in (a) and (c) indicate the expected pulse shape value \TTTave for full-energy neutron interactions. 
		\label{fig:fit_sketches}}
\end{figure*}

Four quantities are calculated to characterize the light emission produced by proton recoil events with known energy and direction within the arbitrarily established set of axes. These are \LOave, \TTTave, \Pave, and \Dave: the expected values of total light, pulse shape parameter, prompt light, and delayed light, respectively. \LOave and \TTTave correspond to the x- and y-coordinates of the endpoint of the upper neutron band in the two-dimensional distribution of $L$~vs.~$S$, shown for one anthracene dataset in~\fig{fig:fit_sketches}~(a). \LOave and \TTTave are calculated from this distribution as follows; the process is repeated for each dataset corresponding to different proton recoil directions in each material.
\nomenclature{\LOave}{Expected total light output}%
\nomenclature{\TTTave}{Expected pulse shape parameter}%
\nomenclature{\Pave}{Expected prompt light output}%
\nomenclature{\Dave}{Expected delayed light output}%

First, neutron events are identified as those events with \TTT above the curved red $n$-$\gamma$ cutoff line and \LO above a light output threshold (\keVee{3000} for \MeV{14.1} measurements and \keVee{300} for \MeV{2.5} measurements). These light output thresholds were set high enough to have good pulse shape separation between the neutron and gamma-ray regions and exclude lower energy contributions such as environmental scatter and signal from carbon recoils. A light output spectrum from the neutron events is produced, as shown in~\fig{fig:fit_sketches}~(b). \LOave and its statistical uncertainty are produced by applying the following fit function:

\begin{equation}\label{eqn:l_fit_fxn}
f(L)=\frac{mL+b}{2}\left[1-\erf\left(\frac{L-\hat{L}}{\sigma\sqrt{2}}\right)\right]-\frac{m\sigma}{\sqrt{2\pi}}e^{\frac{-(L-\hat{L})^2}{2\sigma^2}}
\end{equation}

\eqnref{eqn:l_fit_fxn} models a negatively sloped line with a hard cutoff convolved with a Gaussian resolution function, where \LOave is the expected light output from a full-energy interaction and $\sigma$ is the detector resolution. The resolution $\sigma$ was a floating fit parameter for all materials except DPAC. $\sigma$ was fixed for DPAC in order to ensure consistent fitting despite its poor resolution. 

Neutron events with light output in the range \LOaverange are considered full-energy proton recoil events. This widens the angular selection window by approximately \degrees{15} around the forward direction. 

\TTTave and its statistical uncertainty are calculated as the centroid position of a Gaussian curve fit to the pulse shape distribution of full-energy proton recoil events as shown in \fig{fig:fit_sketches}~(c). These events fall within the two vertical dashed lines in \fig{fig:fit_sketches}~(a).

The expected prompt and delayed light components are calculated from \LOave and \TTTave as shown in Eqns.~\ref{eqn:Pave} and~\ref{eqn:Dave}. These values provide an approximation of the prompt and delayed light by considering everything in the tail region to be delayed light and everything before the delayed region to be prompt light. The samples before the tail region and after the total region, which were included in the calculation of \LOave, were confirmed to have minimal impact on the calculations of \Pave and \Dave. Statistical uncertainties in \Pave and \Dave are propagated from the statistical uncertainties in \LOave and \TTTave. 

\begin{equation}\label{eqn:Pave}
\Pave = \LOave (1-\TTTave)
\end{equation}
\begin{equation}\label{eqn:Dave}
\Dave = \LOave \TTTave 
\end{equation}

This paper expands on previous measurements by the authors in which only the distributions of \LOave and \TTTave were presented~\cite{Schuster2016,Schuster2016a} by including the distributions of the expected prompt and delayed light components. \Pave and \Dave are indicative of the separate behavior of singlet and triplet states, respectively. While \TTTave provides a measurement of their comparative behavior and \LOave a measurement of their combined behavior, \Pave and \Dave provide more detailed information about how the prompt and delayed light emission changes on a fundamental level, as will be discussed in \secref{sec:interpret}.

\section{Results}\label{sec:results}

The scintillation anisotropy measurements are shown in Figs.~\ref{fig:anth_dt} through~\ref{fig:dpac_dt}. Each figure displays the directional distribution of (a)~\LOave, (b)~\TTTave, (c)~\Pave, and (d)~\Dave. The black points indicate the discrete proton recoil directions at which measurements were taken, and the gradient represents a smooth interpolation between those measurements. The colorbar shows the range of values in each distribution with a vertical capped line showing the average statistical uncertainty. The coordinate system is described in~\secref{sec:rot_system}. 

The arbitrary axes for each crystal have been oriented such that the maximum, minimum, and saddle point features in the \LOave distribution are in approximately the same positions. This allows for a quick visual comparison of the relationship between the anisotropies of \LOave, \TTTave, \Pave, and \Dave across materials. A discussion of these relationships follows in \secref{sec:results_qual}.

The magnitude of each anisotropy measurement is expressed as the ratio of the maximum to minimum value in each distribution. For example, $\ALO = \LOave_{\textrm{max}}/\LOave_{\textrm{min}}$. \tab{tab:pure_anis} shows the magnitude of the anisotropies for \LOave, \TTTave, \Pave, and \Dave for all measurements. A discussion of these quantitative results follows in \secref{sec:results_quant}.
\nomenclature{\ALO}{Magnitude of total light output anisotropy}%
\nomenclature{\ATTT}{Magnitude of pulse shape parameter anisotropy}%
\nomenclature{\AP}{Magnitude of prompt light output anisotropy}%
\nomenclature{\AD}{Magnitude of delayed light output anisotropy}%

\subsection{Qualitative}\label{sec:results_qual}

The behavior of the anisotropy effect can be analyzed in terms of the relationship between the  \LOave, \TTTave, \Pave, and \Dave distributions. This relationship will be discussed for all measured materials.

\begin{figure*}[!t]
	\centering
	\subfloat[\LOave (keVee).]{\includegraphics[trim={0cm 0cm 0cm 0cm},clip,width=2.3in]{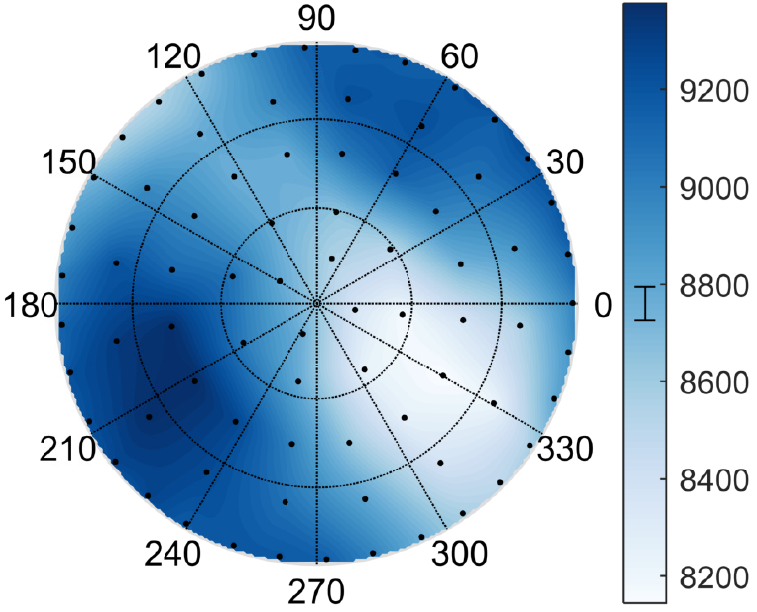}%
		\label{fig:anth_dt_L}}
	\hspace{1em}
	\subfloat[\TTTave.]{\includegraphics[trim={0cm 0cm 0cm 0cm},clip,width=2.3in]{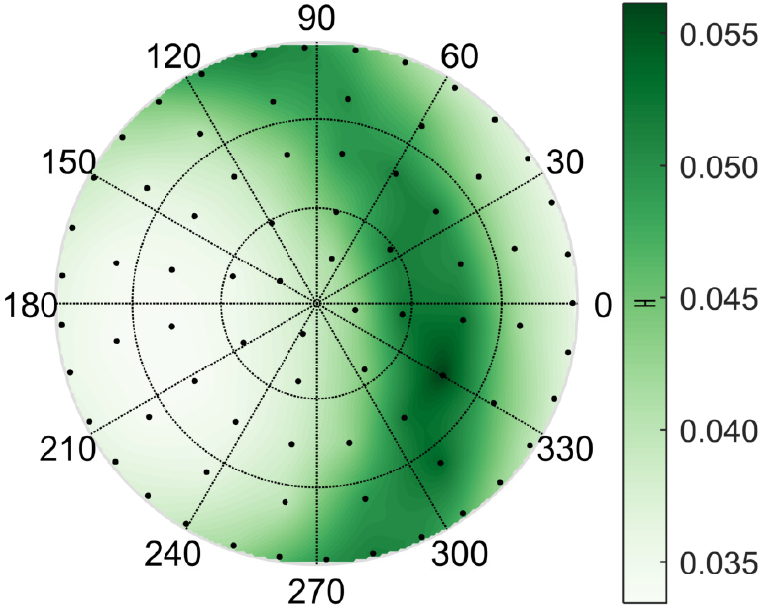}%
		\label{fig:anth_dt_S}}
	
	\subfloat[\Pave (keVee).]{\includegraphics[trim={0cm 0cm 0cm 0cm},clip,width=2.3in]{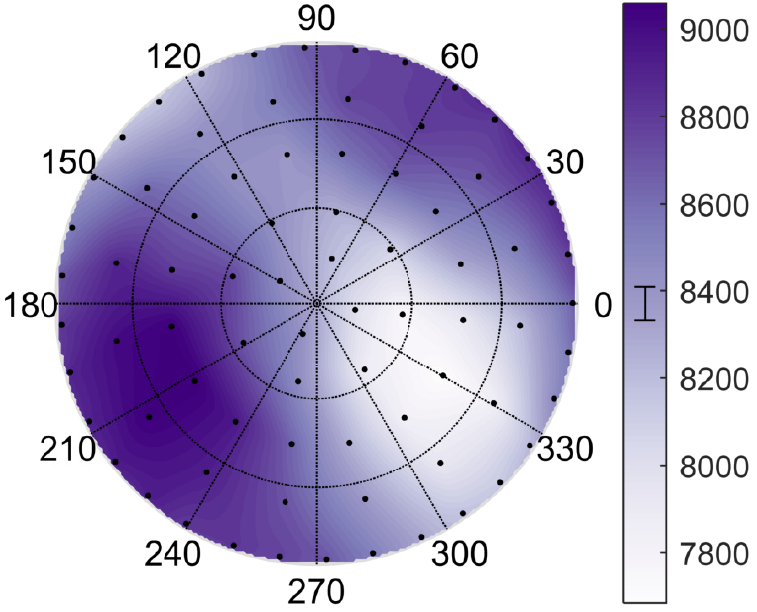}%
		\label{fig:anth_dt_P}}
	\hspace{1em}
	\subfloat[\Dave (keVee).]{\includegraphics[trim={0cm 0cm 0cm 0cm},clip,width=2.3in]{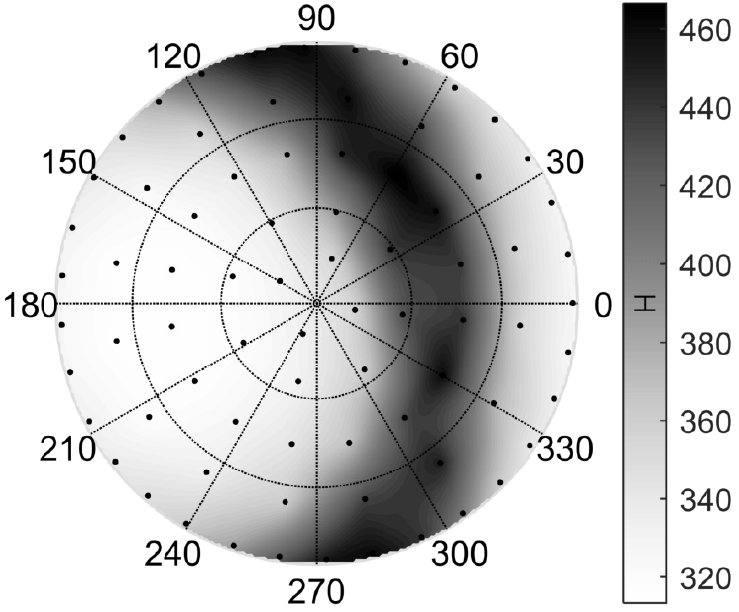}%
		\label{fig:anth_dt_D}}
	\caption{Anthracene; $E_p=$~\MeV{14.1}. \plotsdescribed}
	\label{fig:anth_dt}
\end{figure*}

\begin{figure*}[!t]
	\centering
	\subfloat[\LOave (keVee).]{\includegraphics[trim={0cm 0cm 0cm 0cm},clip,width=2.3in]{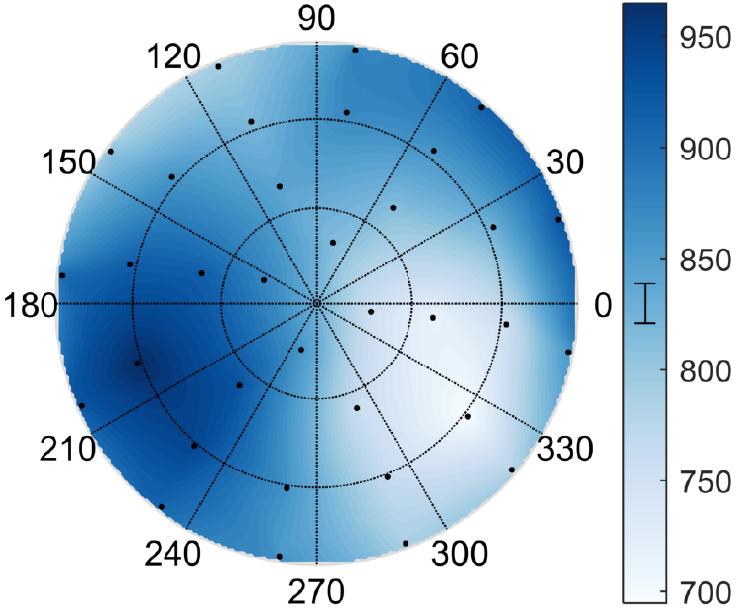}%
		\label{fig:anth_dd_L}}
	\hspace{1em}
	\subfloat[\TTTave.]{\includegraphics[trim={0cm 0cm 0cm 0cm},clip,width=2.3in]{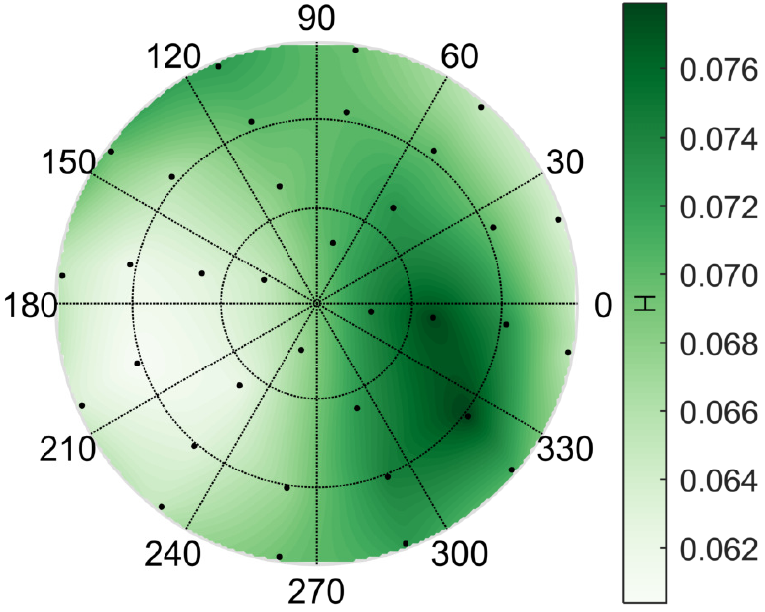}%
		\label{fig:anth_dd_S}}
	
	\subfloat[\Pave (keVee).]{\includegraphics[trim={0cm 0cm 0cm 0cm},clip,width=2.3in]{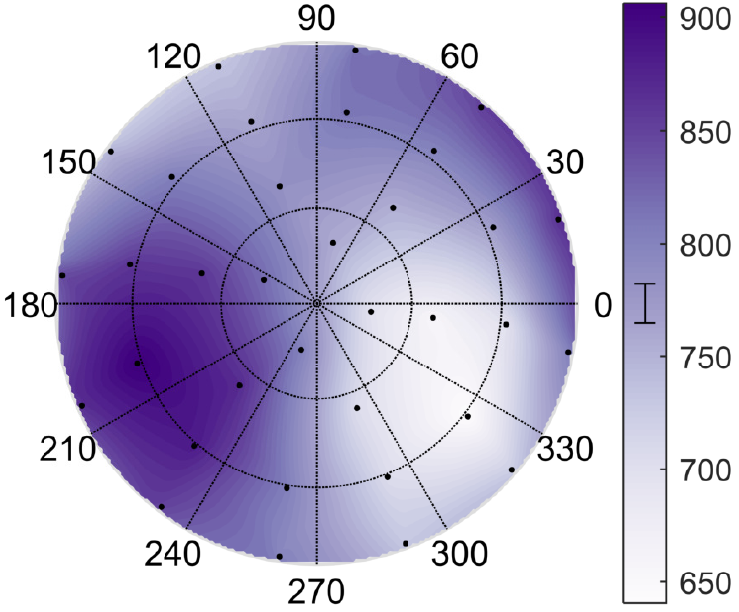}%
		\label{fig:anth_dd_P}}
	\hspace{1em}
	\subfloat[\Dave (keVee).]{\includegraphics[trim={0cm 0cm 0cm 0cm},clip,width=2.3in]{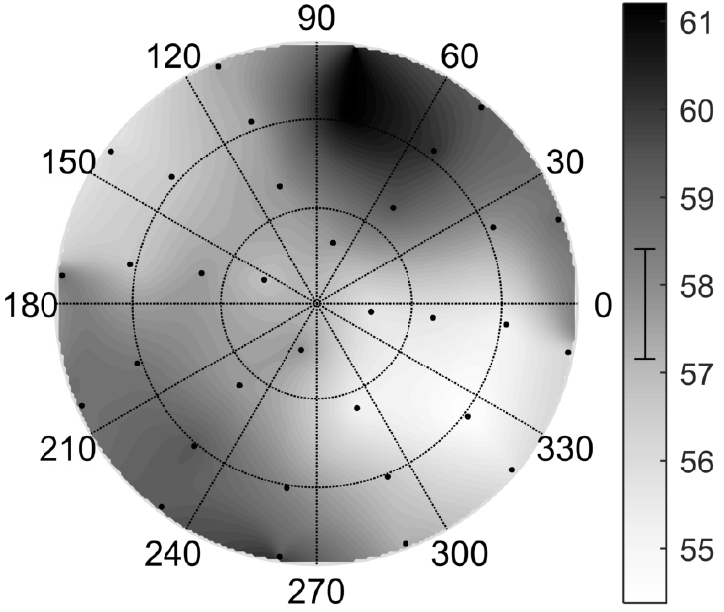}%
		\label{fig:anth_dd_D}}
	\caption{Anthracene; $E_p=$~\MeV{2.5}. \plotsdescribed}
	\label{fig:anth_dd}
\end{figure*}

First, Figs.~\ref{fig:anth_dt} and~\ref{fig:anth_dd} show the scintillation anisotropy in anthracene for \MeV{14.1} and \MeV{2.5} proton recoils, respectively. In both measurements, the positions of the maximum, minimum, and saddle point features in the \LOave and \Pave distributions align, indicating that the prompt scintillation anisotropy dominates the total scintillation anisotropy. At \MeV{14.1}, the \TTTave and \Dave distributions align in their features, indicating that the delayed scintillation anisotropy dominates the pulse shape anisotropy. Interestingly, the position of minimum delayed light moves from approximately $(\theta=60^\circ,\phi=210^\circ)$ to $(\theta=60^\circ,\phi=330^\circ)$ between \MeV{14.1} and \MeV{2.5}. Additionally, while the maximum \Dave direction is the same between \MeV{14.1} and \MeV{2.5}, the maximal \Dave feature appears more like a band at \MeV{14.1} and more like a localized region at \MeV{2.5}. This indicates that the delayed light emission properties may change when the proton recoil crosses a critical energy above or below which the kinetic processes differ significantly. This possibility is also discussed based on quantitative observations in \secref{sec:results_quant}.

Among the four quantities, \Pave and \Dave provide the most fundamental basis to consider, both because they are experimentally independent quantities and because they correspond more closely to distinct physical phenomena, namely the behavior of singlet and triplet excitations. In anthracene, these measurements demonstrate that the maximum and minimum positions of \Pave and \Dave are not aligned, and thus \Pave and \Dave are labeled ``out of sync'' in anthracene.

\begin{figure*}[!t]
	\centering
	\subfloat[\LOave (keVee).]{\includegraphics[trim={0cm 0cm 0cm 0cm},clip,width=2.3in]{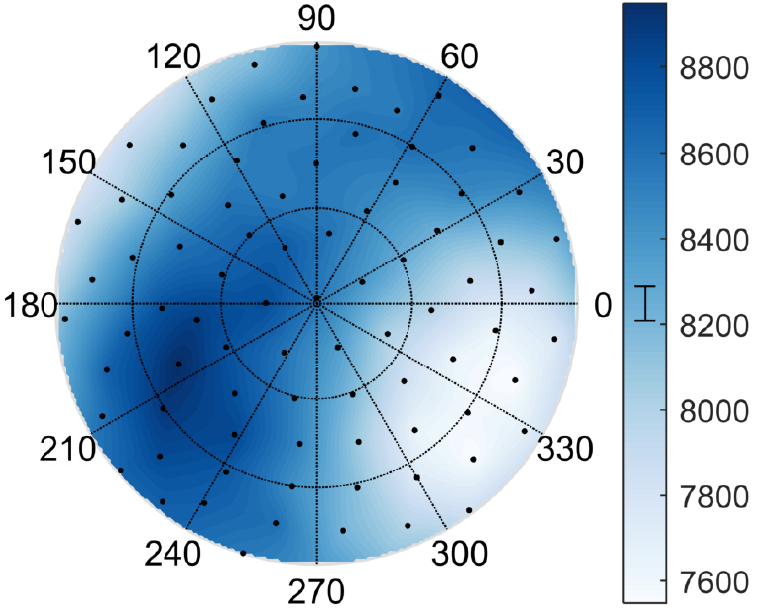}%
		\label{fig:stil_dt_L}}
	\hspace{1em}
	\subfloat[\TTTave.]{\includegraphics[trim={0cm 0cm 0cm 0cm},clip,width=2.3in]{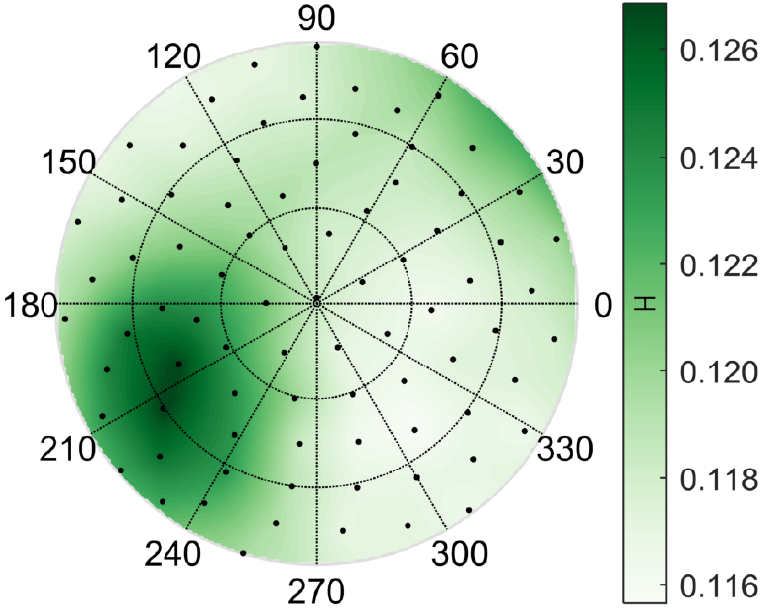}%
		\label{fig:stil_dt_S}}
	
	\subfloat[\Pave (keVee).]{\includegraphics[trim={0cm 0cm 0cm 0cm},clip,width=2.3in]{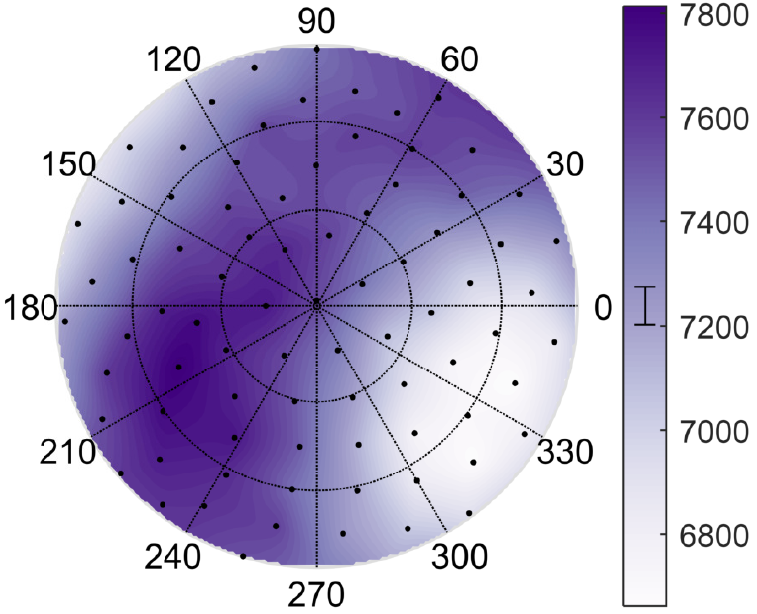}%
		\label{fig:stil_dt_P}}
	\hspace{1em}
	\subfloat[\Dave (keVee).]{\includegraphics[trim={0cm 0cm 0cm 0cm},clip,width=2.3in]{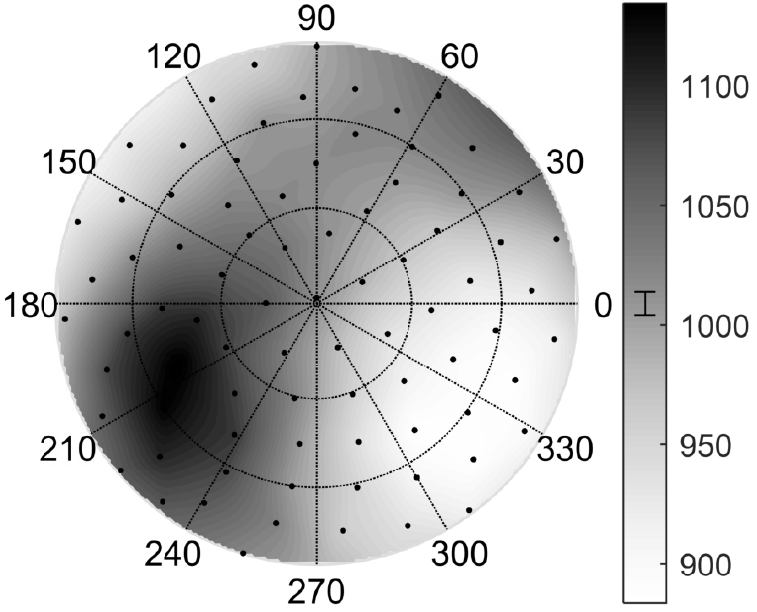}%
		\label{fig:stil_dt_D}}
	\caption{Stilbene; $E_p=$~\MeV{14.1}. \plotsdescribed}
	\label{fig:stil_dt}
\end{figure*}

\begin{figure*}[!t]
	\centering
	\subfloat[\LOave (keVee).]{\includegraphics[trim={0cm 0cm 0cm 0cm},clip,width=2.3in]{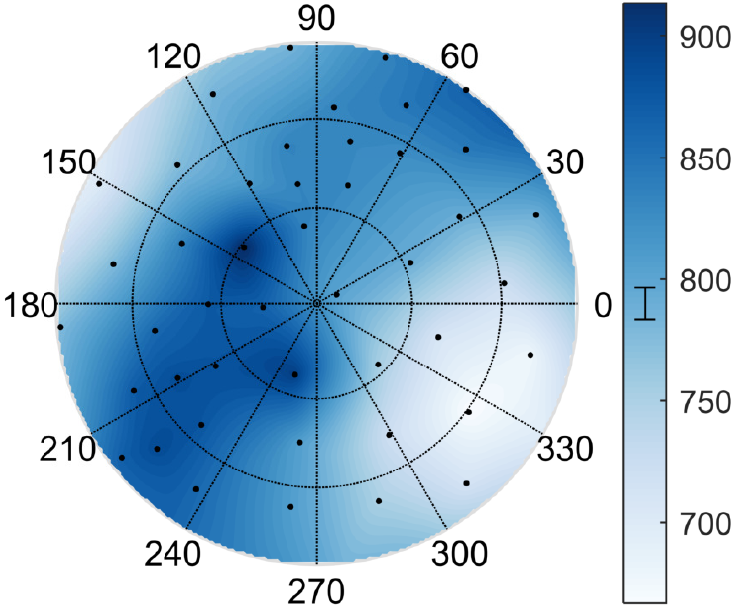}%
		\label{fig:stil_dd_L}}
	\hspace{1em}
	\subfloat[\TTTave.]{\includegraphics[trim={0cm 0cm 0cm 0cm},clip,width=2.3in]{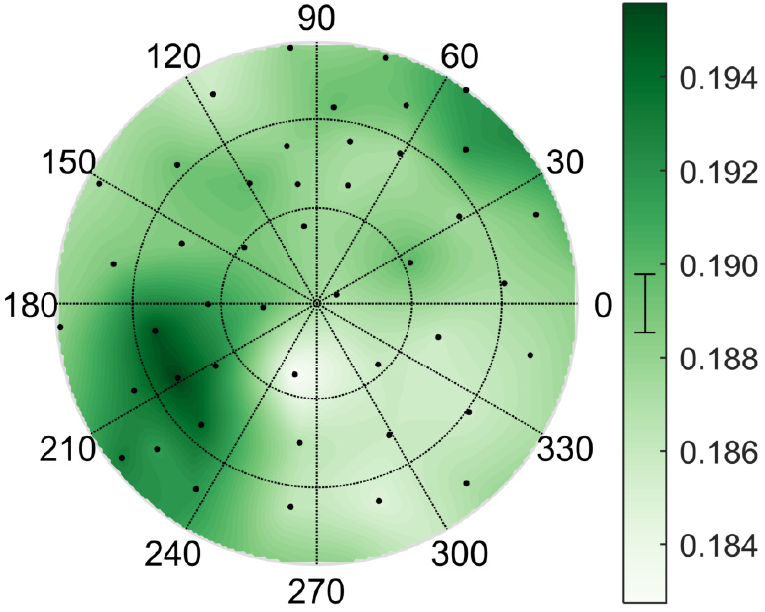}%
		\label{fig:stil_dd_S}}
	
	\subfloat[\Pave (keVee).]{\includegraphics[trim={0cm 0cm 0cm 0cm},clip,width=2.3in]{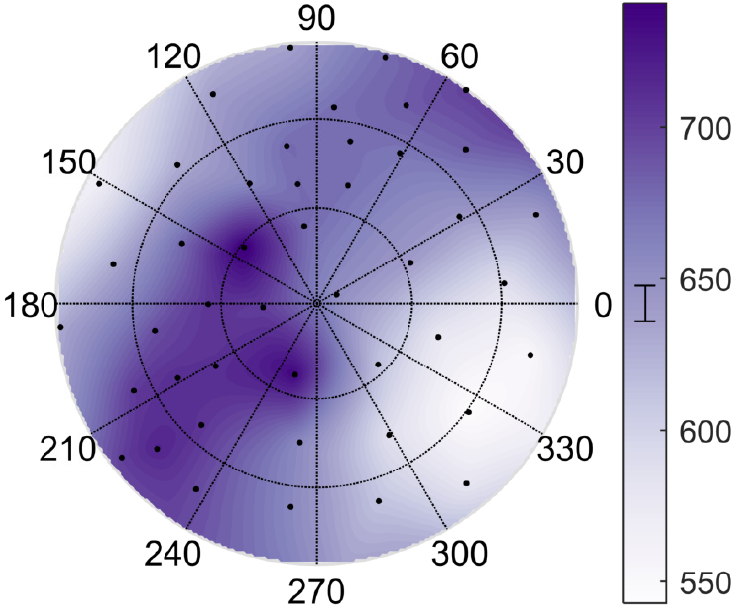}%
		\label{fig:stil_dd_P}}
	\hspace{1em}
	\subfloat[\Dave (keVee).]{\includegraphics[trim={0cm 0cm 0cm 0cm},clip,width=2.3in]{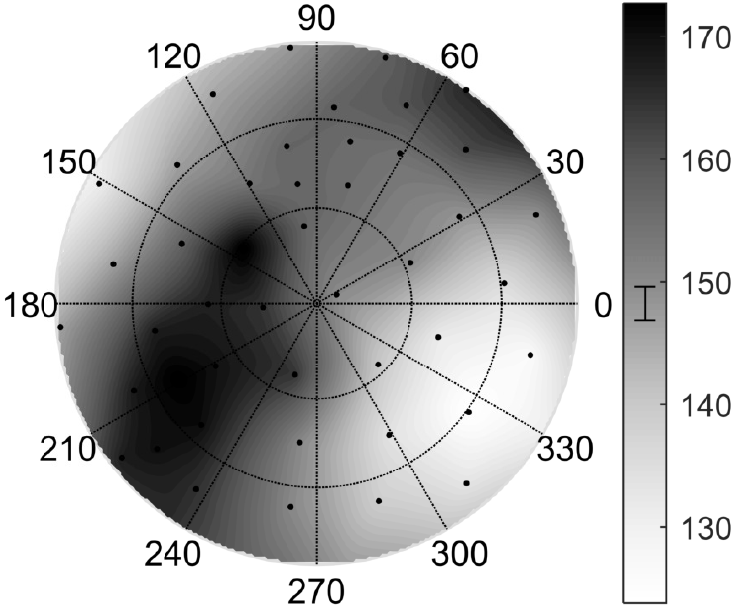}%
		\label{fig:stil_dd_D}}
	\caption{Stilbene; $E_p=$~\MeV{2.5}. \plotsdescribed}
	\label{fig:stil_dd}
\end{figure*}

Next, Figs.~\ref{fig:stil_dt} and~\ref{fig:stil_dd} show the scintillation anisotropy in stilbene for \MeV{14.1} and \MeV{2.5} proton recoils, respectively. In both measurements, the positions of the maximum, minimum, and saddle point are aligned for all four metrics. Thus, the behaviors of \Pave and \Dave in stilbene at \MeV{14.1} and \MeV{2.5} are ``in sync.'' 

\begin{figure*}[!t]
	\centering
	\subfloat[\LOave (keVee).]{\includegraphics[trim={0cm 0cm 0cm 0cm},clip,width=2.3in]{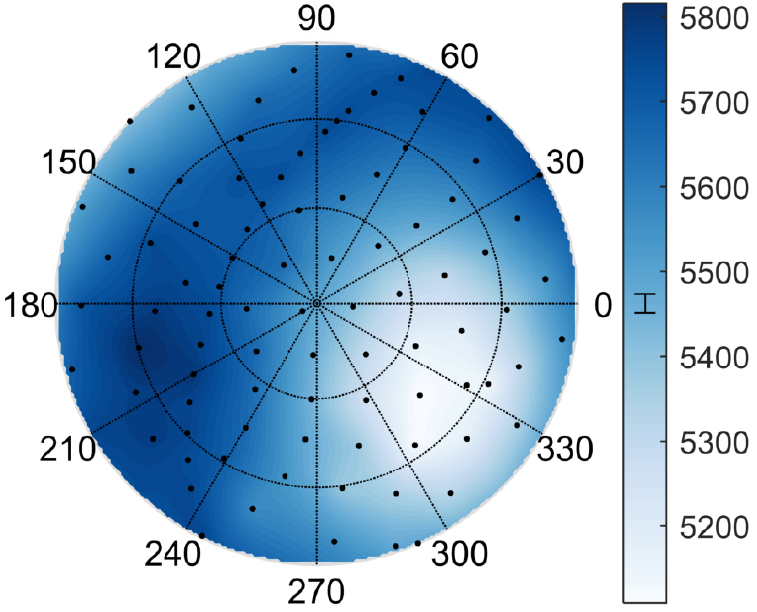}%
		\label{fig:pter_dt_L}}
	\hspace{1em}
	\subfloat[\TTTave.]{\includegraphics[trim={0cm 0cm 0cm 0cm},clip,width=2.3in]{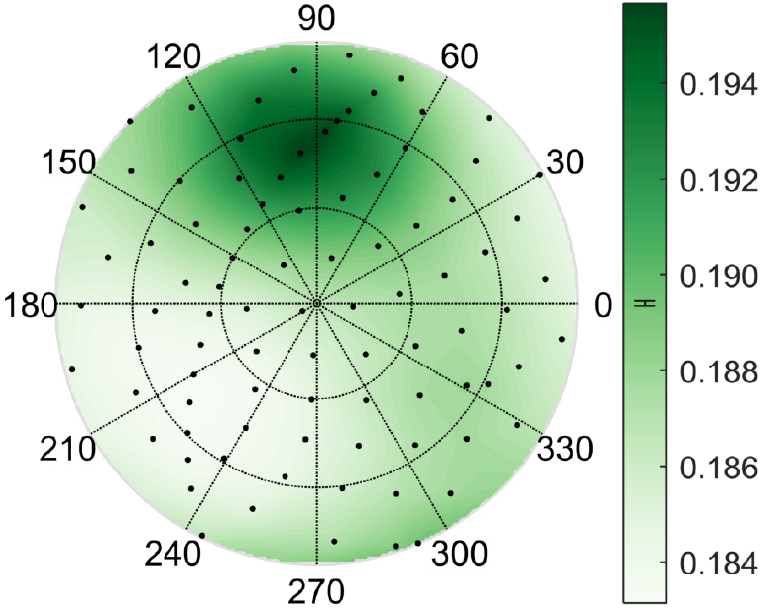}%
		\label{fig:pter_dt_S}}
	
	\subfloat[\Pave (keVee).]{\includegraphics[trim={0cm 0cm 0cm 0cm},clip,width=2.3in]{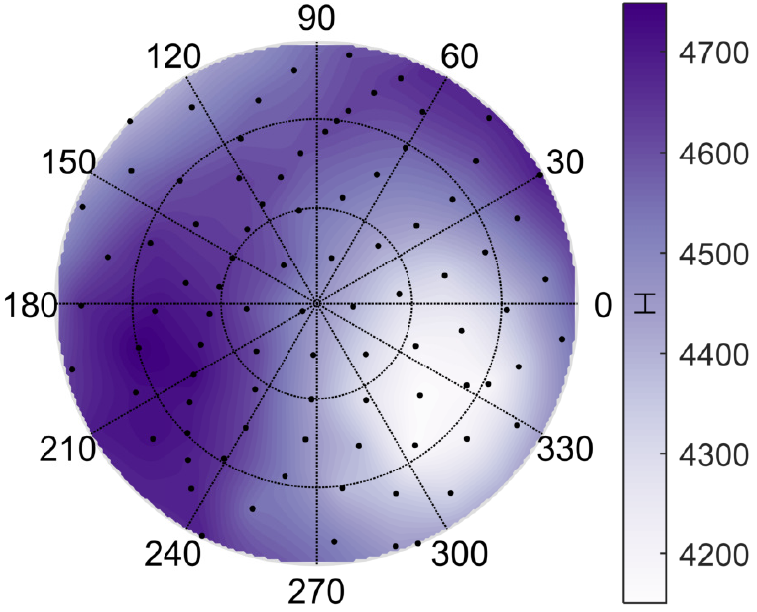}%
		\label{fig:pter_dt_P}}
	\hspace{1em}
	\subfloat[\Dave (keVee).]{\includegraphics[trim={0cm 0cm 0cm 0cm},clip,width=2.3in]{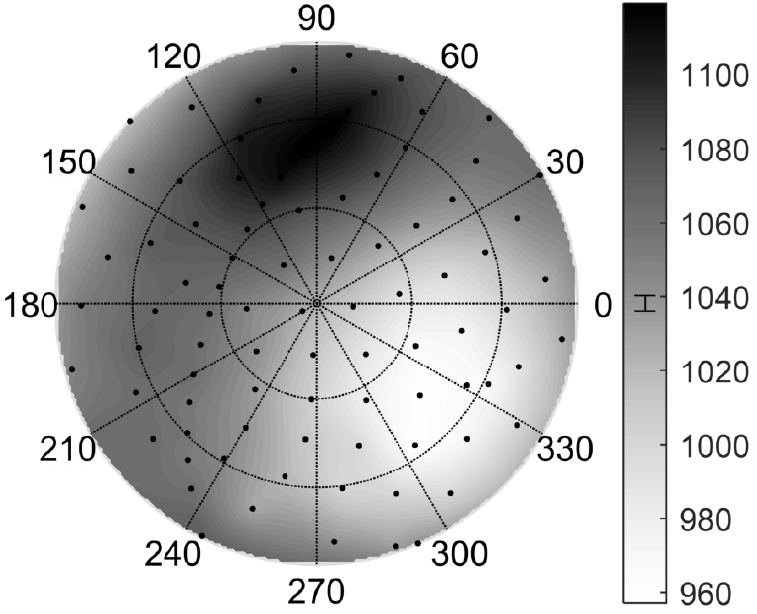}%
		\label{fig:pter_dt_D}}
	\caption{P-terphenyl; $E_p=$~\MeV{14.1}. \plotsdescribed}
	\label{fig:pter_dt}
\end{figure*}

Continuing on, \fig{fig:pter_dt} shows measurements for \MeV{14.1} proton recoils on p-terphenyl. Similar to anthracene, \Pave and \Dave are out of sync. Also similar to anthracene, the behavior of \LOave follows that of \Pave, and the behavior of \TTTave follows that of \Dave. Thus, it appears that the total light anisotropy is dominated by the prompt light anisotropy, and the pulse shape anisotropy is dominated by the delayed light anisotropy. 

\begin{figure*}[!t]
	\centering
	\subfloat[\LOave (keVee).]{\includegraphics[trim={0cm 0cm 0cm 0cm},clip,width=2.3in]{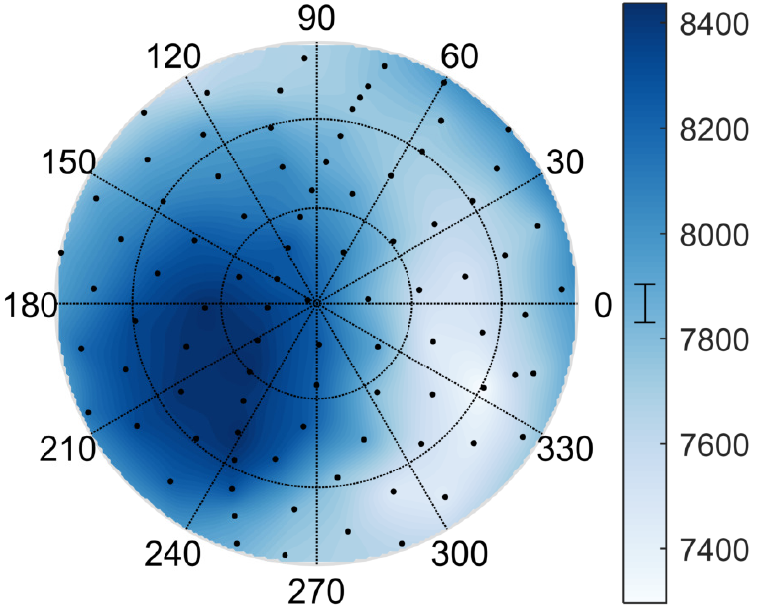}%
		\label{fig:bb_dt_L}}
	\hspace{1em}
	\subfloat[\TTTave.]{\includegraphics[trim={0cm 0cm 0cm 0cm},clip,width=2.3in]{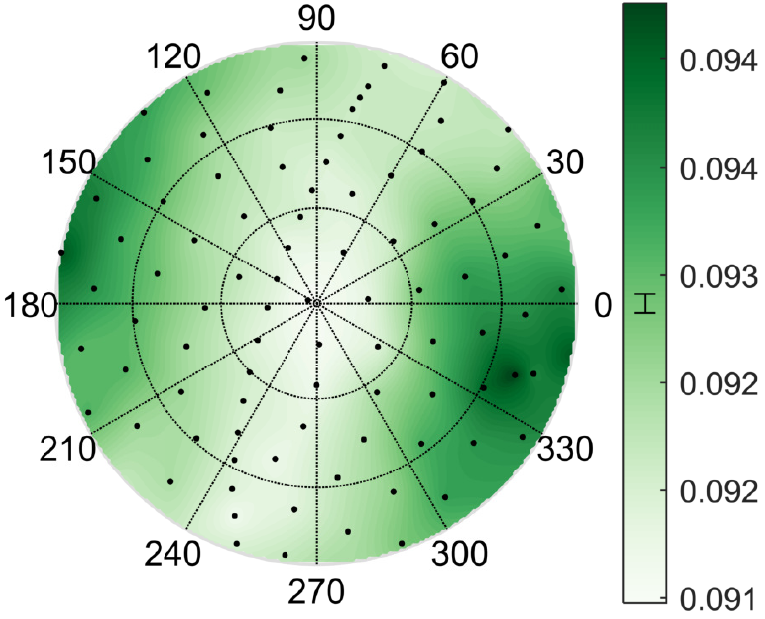}%
		\label{fig:bb_dt_S}}
	
	\subfloat[\Pave (keVee).]{\includegraphics[trim={0cm 0cm 0cm 0cm},clip,width=2.3in]{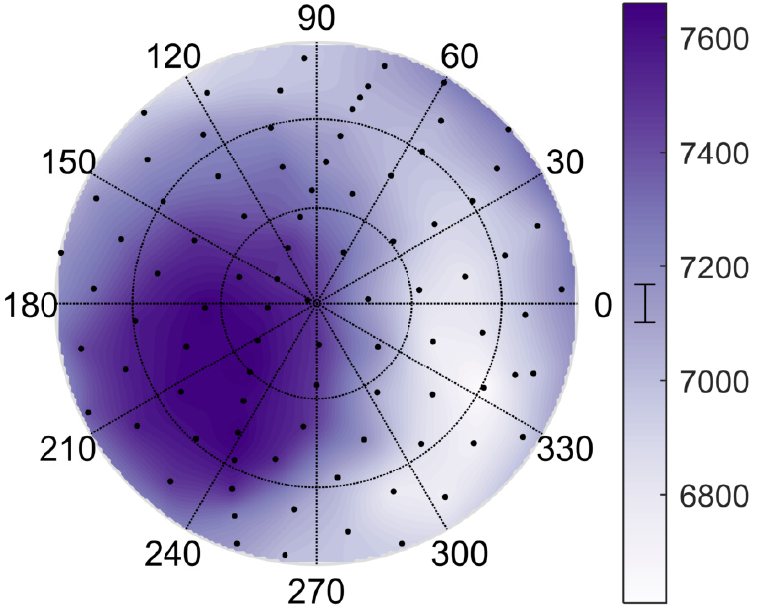}%
		\label{fig:bb_dt_P}}
	\hspace{1em}
	\subfloat[\Dave (keVee).]{\includegraphics[trim={0cm 0cm 0cm 0cm},clip,width=2.3in]{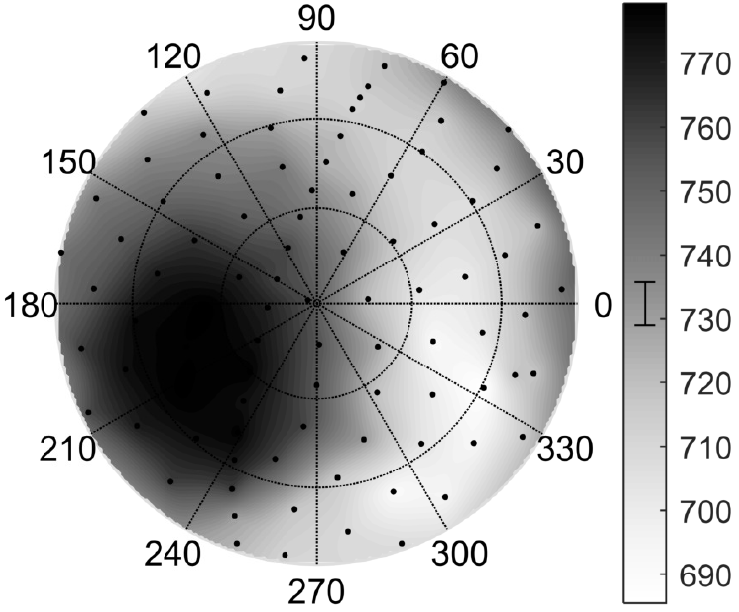}%
		\label{fig:bb_dt_D}}
	\caption{Bibenzyl; $E_p=$~\MeV{14.1}. \plotsdescribed}
	\label{fig:bb_dt}
\end{figure*}

\begin{figure*}[!t]
	\centering
	\subfloat[\LOave (keVee).]{\includegraphics[trim={0cm 0cm 0cm 0cm},clip,width=2.3in]{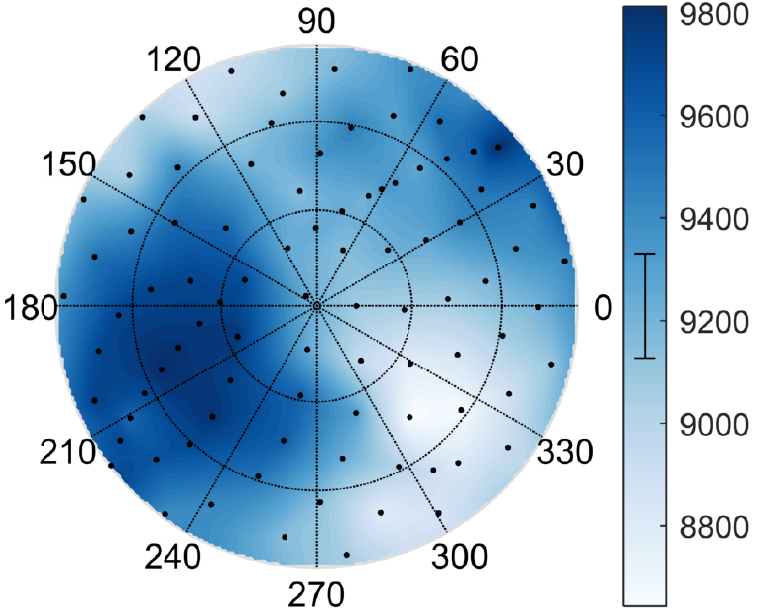}%
		\label{fig:dpac_dt_L}}
	\hspace{1em}
	\subfloat[\TTTave.]{\includegraphics[trim={0cm 0cm 0cm 0cm},clip,width=2.3in]{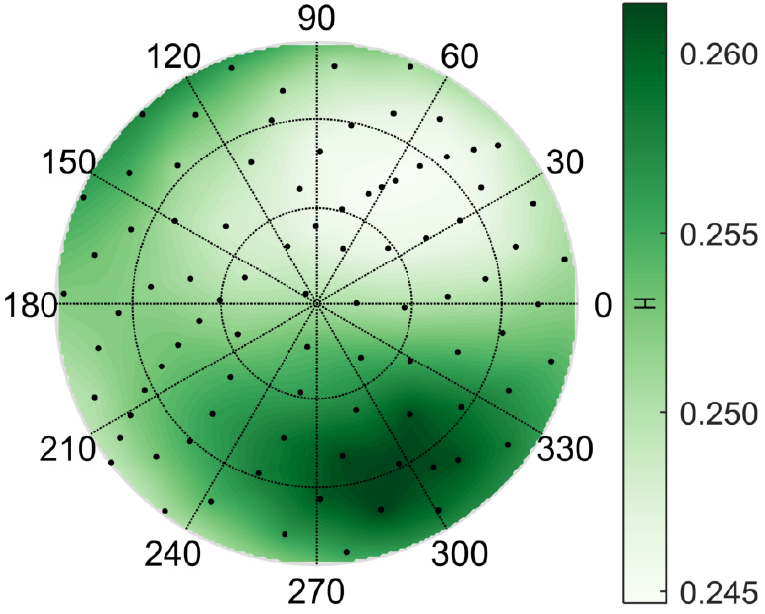}%
		\label{fig:dpac_dt_S}}
	
	\subfloat[\Pave (keVee).]{\includegraphics[trim={0cm 0cm 0cm 0cm},clip,width=2.3in]{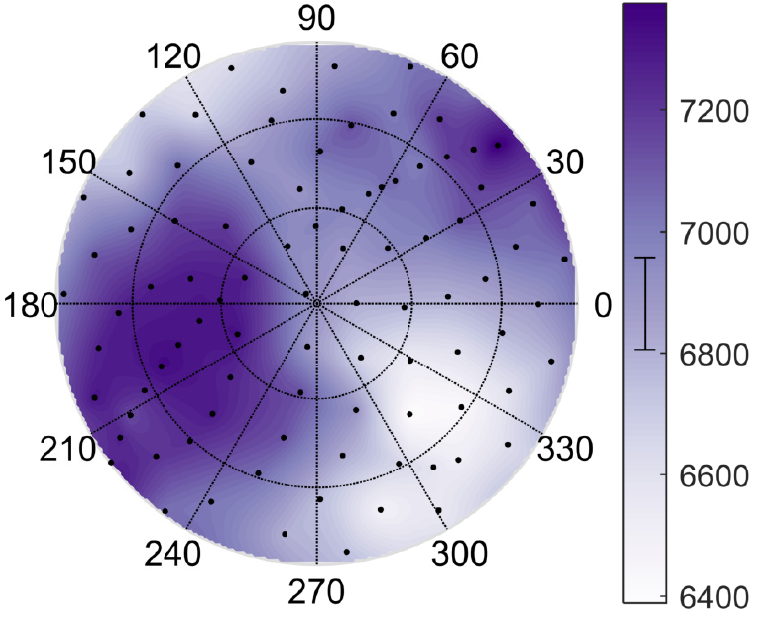}%
		\label{fig:dpac_dt_P}}
	\hspace{1em}
	\subfloat[\Dave (keVee).]{\includegraphics[trim={0cm 0cm 0cm 0cm},clip,width=2.3in]{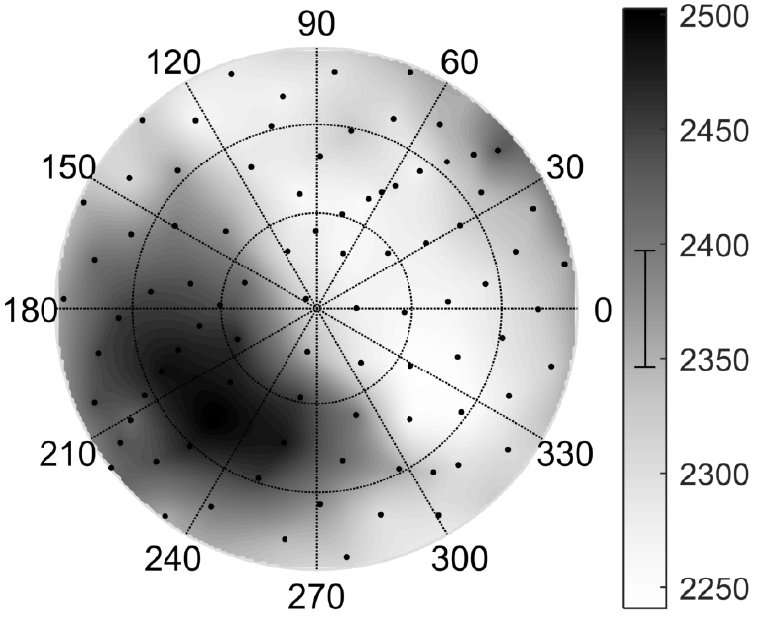}%
		\label{fig:dpac_dt_D}}
	\caption{Diphenylacetylene; $E_p=$~\MeV{14.1}. \plotsdescribed}
	\label{fig:dpac_dt}
\end{figure*}

Finally, Figs.~\ref{fig:bb_dt} and~\ref{fig:dpac_dt} show measurements of \MeV{14.1} proton recoils on bibenzyl and diphenylacetylene (DPAC), respectively. Both of these materials' effects follow that of stilbene in that the prompt and delayed light components' behaviors are in sync. Interestingly, the pulse shape anisotropies in these materials both differ from the behavior of \LOave, \Pave, and \Dave. These measurements show that the pulse shape anisotropy can be misleading in understanding how the light emission changes with angle. Previously, when these results were interpreted only in terms of \LOave and \TTTave, it appeared that bibenzyl and DPAC had significantly unusual behavior in their scintillation anisotropies because the \TTTave behavior was so different than that of \LOave. Now it can be seen that, more fundamentally, the relationships between \Pave and \Dave are in sync, so the features in \TTTave reflect differences in the contours of \Pave and \Dave, not the maximal or minimal directions for \Pave and \Dave. Thus, the results from bibenzyl and DPAC provide examples of why it is important to analyze \Pave and \Dave in addition to \LOave and \TTTave.

To conclude the qualitative observations of the behavior of the scintillation anisotropy, these results demonstrate that the total light output \LOave has the same behavior as the prompt light emission \Pave in all materials, while the behavior of \TTTave and \Dave may vary. In anthracene and p-terphenyl, \Pave and \Dave are out of sync. In stilbene, bibenzyl, and diphenylacetylene, \Pave and \Dave are in sync. These relationships will be discussed in terms of the crystal structures in \secref{sec:interpret}.

\subsection{Quantitative}\label{sec:results_quant}

The anisotropy measurements can also be analyzed in terms of the magnitude of the effect, calculated as the ratio of maximum to minimum measurement, $A$, for each quantity. These values are shown in \tab{tab:pure_anis}.

\begin{table*}[htbp]
	\centering
	\caption{Magnitude of change in \LOave, \TTTave, \Pave, and \Dave values measured for \MeV{14.1} (unshaded) and \MeV{2.5} (shaded) proton recoil events in various pure crystal detectors, and the corresponding figure number. Uncertainties are statistical only.}
	\label{tab:pure_anis}
	\begin{tabular}{c|c|l|l|l|l|c}
		                      & $E_p$ (MeV) & \ALO       & \ATTT    & \AP       &  \AD & Fig. \\
		\hline       
		\multirow{2}{*}{Anthracene} & 14.1  & 1.155(5)   & 1.700(6) & 1.183(7)  & 1.502(9)  & \ref{fig:anth_dt}\\ 
		                            & \shd 2.5 & \shd 1.398(31)  & \shd 1.297(5) & \shd 1.425(32) & \shd 1.129(22) & \shd \ref{fig:anth_dd}\\ 
		\multirow{2}{*}{Stilbene}   & 14.1  & 1.191(8)   & 1.100(1) & 1.178(8)  & 1.291(9)  & \ref{fig:stil_dt}\\
		                            & \shd 2.5   & \shd 1.382(23)  & \shd 1.073(17)& \shd 1.377(30) & \shd 1.408(30) & \shd \ref{fig:stil_dd} \\
		P-terphenyl                 & 14.1  & 1.142(3)   & 1.070(1) & 1.147(3)  & 1.173(3)  & \ref{fig:pter_dt}\\
		
		Bibenzyl                    & 14.1  & 1.161(8)   & 1.032(1) & 1.164(8)  & 1.141(6)  & \ref{fig:bb_dt}  \\
		Diphenylacetylene           & 14.1  & 1.139(20)  & 1.070(1) & 1.160(27) & 1.120(14) & \ref{fig:dpac_dt}
	\end{tabular}%
\end{table*}%

There are several interesting conclusions that can be drawn from considering the $A$ values in \tab{tab:pure_anis}. First, consider the dependence on proton recoil energy. Measurements on anthracene and stilbene were made at two energies: \MeV{14.1} and \MeV{2.5}. In both materials, \ALO is negatively correlated with energy, meaning that \ALO is smaller at the higher energy, while \ATTT is positively correlated with energy. While this was already demonstrated in previous work, prior analyses do not provide insight on whether the changes are due to the prompt or delayed light components~\cite{Schuster2016,Schuster2016a}. The new analysis of the prompt and delayed components in this work adds new information. In both materials, the prompt anisotropy, \AP, is negatively correlated with energy. Interestingly, the delayed light anisotropy, \AD, is negatively correlated with energy in stilbene and positively correlated with energy in anthracene. This result adds to questions surrounding the delayed light emission in anthracene. 

The delayed light anisotropy in anthracene at \MeV{14.1}~vs.~\MeV{2.5} has several interesting features. As discussed in \secref{sec:results_qual}, the direction of the minimum of the delayed light emission changes between \MeV{14.1} and \MeV{2.5}, while it is the same in stilbene at both energies. Additionally, \AD changes significantly in magnitude from $1.502(9)$ at \MeV{14.1} to $1.129(22)$ at \MeV{2.5}. The measurement of \AD at \MeV{14.1} conflicts with prior authors' conclusions that the delayed light anisotropy is minimal in anthracene~\cite{Tsukada1962,Brooks1974}. \citeauthor{Tsukada1962} measured a small \AD at \MeV{3.7}~\cite{Tsukada1962}, while \citeauthor{Brooks1974} did not measure \AD directly but applied the conclusion in~\cite{Tsukada1962} to his measurements from \MeV{1} to \MeV{22}, attributing a large \ATTT value to changes in the prompt emission, not the delayed emission~\cite{Brooks1974}. The measurements presented in this paper demonstrate that \AD is small at \MeV{2.5} but very large at \MeV{14.1}. The change in magnitude and behavior between \MeV{14.1}~vs.~3.7 and \MeV{2.5} may indicate that the delayed light emission properties change significantly when the proton recoil crosses a critical energy somewhere between \MeV{14.1} and \MeV{3.7}. 


Next, one can compare the magnitude of the anisotropy effect across all materials at \MeV{14.1}. The total and prompt light anisotropies, \ALO and \AP, are on approximately the same order across all materials. This, together with the observation in \secref{sec:results_qual} that the behavior of \LOave and \Pave are always in sync, supports a theory that \Pave dominates the quantitative and qualitative effects in \LOave, and that these effects are roughly uniform across all materials.

Lastly, the \ATTT values reflect whether \AP and \AD are in or out of sync in each material. \ATTT is much larger in anthracene than any other materials, partly because \AD is much larger in anthracene, but also because the prompt and delayed distributions are out of sync, so \ATTT amplifies their differences. \ATTT in anthracene is approximately the product of \ALO and \AD, not the ratio as it is in other materials. By comparison, bibenzyl, whose prompt and delayed emissions are in sync, has the smallest \ATTT value despite having \ALO, \AP, and \AD values comparable to many other materials.

\subsection{Systematic errors}

The error bars in \tab{tab:pure_anis} and on the colorbars for Figs.~\ref{fig:anth_dt} through~\ref{fig:dpac_dt} represent the statistical uncertainties due to counting statistics. Other systematic errors exist in the measurements that are not quantified. These include the following:

The scintillation response in organic scintillator materials has been shown to vary with temperature~\cite{Schuster2016}, including the magnitude of the anisotropy \ALO~\cite{Heckmann1961}. Temperature effects are not expected to be significant for these measurements because the largest temperature change in any measurement was \degreesC{2.1}. Additionally, distributions of \LOave, \TTTave, \Pave, and \Dave~vs.~temperature were all viewed and did not demonstrate any strong correlation, some of which were included in~\cite{SchusterThesis2016}. 

While each measurement is made at a fixed angle between the neutron generator and the detector, there is some width to the range of proton recoil events selected. First, the incident neutron direction has a width of up to \degrees{1.2} due to the size of the detector. Second, the selection of events within the light output range \LOaverange widens the proton recoil range to approximately within \degrees{15} of the forward direction. This means that the values \LOave, \TTTave, \Pave, and \Dave are averages over a small cone of proton recoil directions. Thus, the $A$ values in \tab{tab:pure_anis} are biased toward smaller values because measurements at the maximal and minimal directions include proton recoils within a small region around the ``true'' maximal and minimal light output directions.

\section{Interpretation}\label{sec:interpret}

\subsection{Basics of light emission}\label{sec:lightemission}

In order to understand why the scintillation anisotropy occurs in these crystals, one must first understand the fundamentals of light emission in organic scintillator materials. A brief description follows; more thorough descriptions can be found in~\cite{Birks1964,Cowan1976,Brooks1979}.

\begin{figure}
	\centering
	\includegraphics[trim={0cm 0cm 0cm 0cm},clip,width=.35\textwidth]{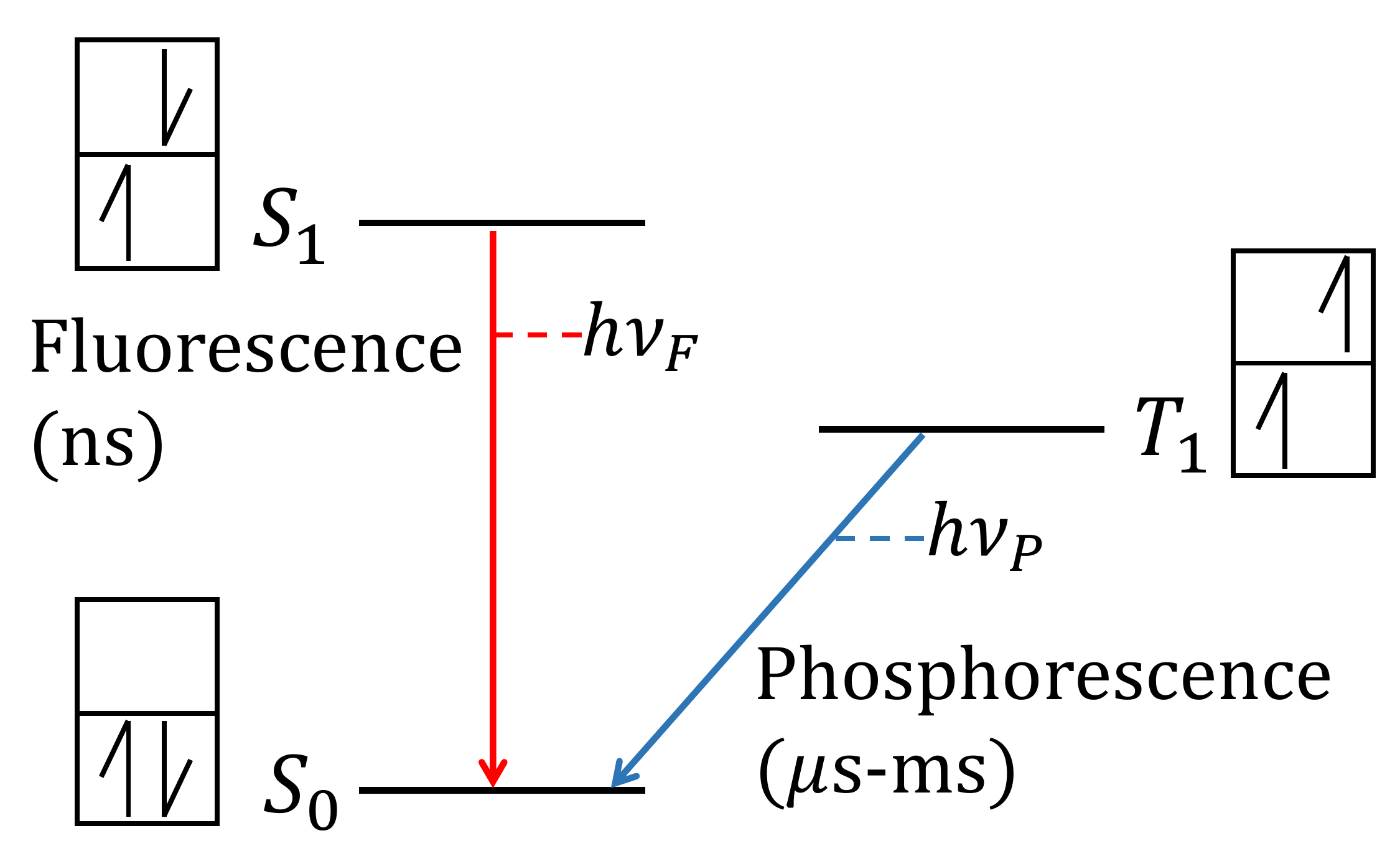}
	\caption{Simplified Jablonski diagram showing the first electronic excited singlet and triplet states and the ground singlet state.
		\label{fig:Jablonski}
	}
\end{figure}

Some of the energy deposited by radiation in an organic crystal will produce molecular excitations of the $\pi$-molecular orbitals, often given particle-like status and referred to as ``excitons.'' On a very short time scale, these excitations relax from higher states into the first electronic excited singlet state, \Sone, or triplet state, \Tone. Ground state molecules are in a singlet ground state, \Sgrd. \fig{fig:Jablonski} shows these states on a simplified Jablonski diagram, which organizes excited states vertically by excitation energy and horizontally by multiplicity. These $\pi$-orbital excitations are responsible for light emission and are, therefore, important to understanding the scintillation anisotropy effect. 
\nomenclature{\Sgrd}{Singlet ground state}%
\nomenclature{\Sone}{First electronic excited singlet state}%
\nomenclature{\Tone}{First electronic excited triplet state}

\begin{figure}
	\centering
	\includegraphics[trim={0cm 0cm 0cm 0cm},clip,width=.32\textwidth]{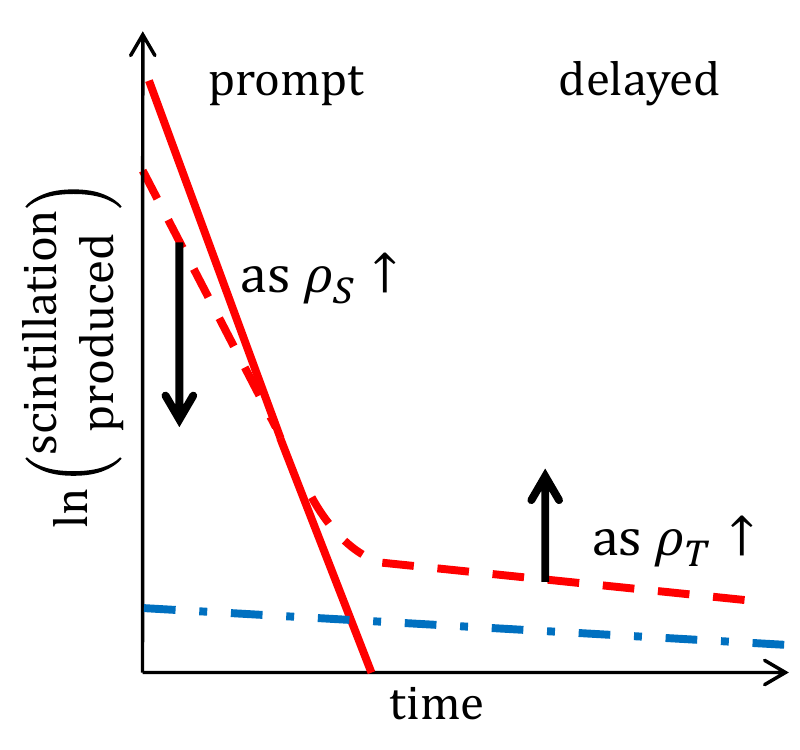}
	\caption{Illustration of light emission from major kinetic processes. The solid single red line is the prompt light emission from fluorescence. The dot-dashed blue line is phosphorence which is generally not observed in any significant amount. The dashed red line in the prompt time region indicates the reduced prompt light emission as a result of SIQ, and the dashed red line in the delayed time region indicates increased delayed light emission from TTA. 
	\label{fig:scint_trace}}
\end{figure}

Many kinetic processes are available to these excitations. Five processes that are especially important to understanding the anisotropy effect will be discussed here. The first process is fluorescence, during which a singlet excitation decays to ground state and emits a photon, as represented by the red arrow in~\fig{fig:Jablonski}. This process occurs rapidly, on the nanosecond time scale. The fluorescence of the singlet excitations from the initial population is often referred to as the ``prompt'' light emission. This contribution to the scintillation is illustrated as the solid red line in~\fig{fig:scint_trace}.

The second process is phosphorescence, during which a triplet excitation decays to ground state and emits a photon. This process is spin-forbidden, and, as a result, has a much longer time scale of $\mu$s to ms. This process occurs much slower than the time scale of our measurements, and thus phosphorescence is not observed in any significant amount in our measurements. The contribution from phosphorescent emissions is illustrated as the dot-dashed blue line in~\fig{fig:scint_trace}, but one may proceed as if it does not occur due to its small contribution compared to other processes. 

The third and fourth processes require interactions between two excitations. The third process is singlet ionization quenching (SIQ)~\cite{Birks1964}, in which two singlet states interact as follows:

$$\Sone + \Sone \rightarrow S^*_n + \Sgrd. $$

After the interaction, one of the original singlet excitations may be promoted to a higher-order excitation level. It will then relax to \Sone, and may then de-excite by fluorescence. The other original singlet excitation is left in ground state. The term ``quench'' means any loss of light from the system and is applied here because only one photon is produced from two excitations. Thus, this process decreases the amount of prompt light produced by the original singlet population when the spatial density of singlet excitations, \rhoS, is higher, as indicated by the reduction in the prompt light in~\fig{fig:scint_trace}.
\nomenclature{\rhoS}{Spatial density of singlet excitations}

The fourth process is called triplet-triplet annihilation (TTA)~\cite{Azumi1963}. In this process, two triplets annihilate with one another as follows:

$$\Tone + \Tone \rightarrow \Sone + \Sgrd. $$

The \Sone excitation may then emit light via fluorescence. This process provides a mechanism for triplet excitation energy to be observed as fluorescence on a shorter time scale than phosphorescence. This time scale is generally longer than the prompt singlet fluorescence, and light emitted following TTA is thus referred to as the ``delayed'' light emission. Since the process requires an interaction between excitations, the time scale on which that delayed light is produced depends on the rate at which triplet states meet, governed by their spatial density, \rhoT, and mobilities. The arrow on the right side of \fig{fig:scint_trace} shows how the delayed light increases due to increased TTA when the triplet density, \rhoT, is higher.
\nomenclature{\rhoT}{Spatial density of triplet excitations}

A fifth process that participates is excitation transport. Transport changes the excitation density over the lifetime of the excitations, and therefore changes the rates of SIQ and TTA. The effects of exciton transport in the anisotropy will be discussed more in \secref{sec:hyp_anis}.

\subsection{Hypothesis behind scintillation anisotropy effect}\label{sec:hyp_anis}

The total amount of light emitted and its time distribution depend on the relative amounts of the kinetic processes described in \secref{sec:lightemission}. Since each excitation has multiple kinetic processes available to it, there is competition between them. Because the rates of SIQ and TTA depend on the distance between the two excitations that will participate in the process, the relative rates of kinetic processes will vary as the density of excited states changes. This explains why the pulse shapes produced by neutron and gamma-ray interactions differ: Neutrons produce a proton recoil which deposits energy with a higher \dEdx than the electron recoil produced by the gamma-ray interaction. Thus, neutrons populate a higher density of excitations, which leads to more SIQ and TTA. This results in an event with less prompt light and more delayed light, which is measured as a higher ``tail-to-total'' value in the common charge integration pulse shape discrimination method~\cite{Brooks1959,Adams1978}.

The scintillation anisotropy is believed to be due to a number of kinetic processes that are directionally dependent for excitation distributions within a crystalline material~\cite{Tsukada1965,Brooks1974}. Any kinetic process that involves interactions between excitations may have a preferred interaction orientation based on the physical mechanism by which the interaction occurs. In crystalline materials, the molecules have a fixed, repeating pattern in their orientation and distance to each neighbor. This structure means that several kinetic processes may have preferred directions for interaction within the crystal axes. These include three discussed in \secref{sec:lightemission}: SIQ, TTA, and excitation transport.

\begin{figure*}
	\centering
	\includegraphics[trim={0cm 0cm 0cm 0cm},clip,width=.75\textwidth]{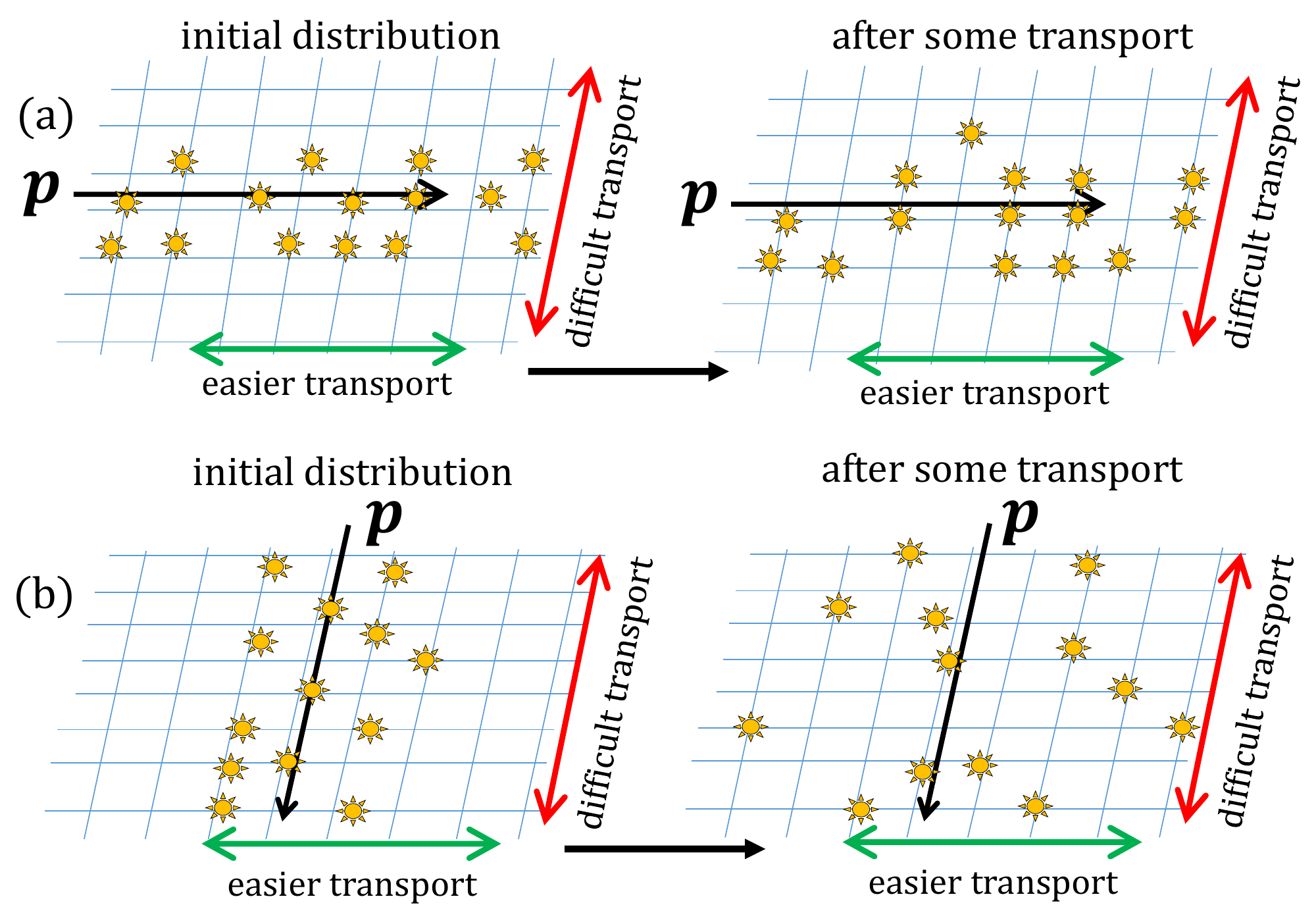}
	\caption{Illustration of excitation transport over time when excitations are populated by a proton recoil $p$ along crystal axes with (a) easy transport or (b) difficult transport. In each, the left cartoon illustrates the initial excitation distribution and the right cartoon illustrates the distribution after some time has passed and excitation transport has occurred. In (a), the density of excitations remains high over time as the excitations move within their initial spatial distribution, while in (b), the density of excitations drops off over time as excitations move away from their initial spatial distribution. In (a), one would observe higher rates of SIQ and TTA over time, leading to a lower \Pave and a higher \Dave. 
		\label{fig:anis_cartoon}}
\end{figure*}

If one simply considers the directionality in excitation transport, it is possible to explain how anisotropy can exist on an elementary level. \fig{fig:anis_cartoon} shows an illustration of excitation transport within a two-dimensional lattice with directions of easy and difficult transport. In \fig{fig:anis_cartoon}~(a), excitations are deposited along an axis with easy excitation transport, and the excitations move within the region of their initial distribution over time so that the density of excitations remains high. In \fig{fig:anis_cartoon}~(b), excitations are deposited along a direction with difficult transport, so excitations move away from their initial distribution over time, producing a lower excitation density.  

\begin{figure*}
	\centering
	\includegraphics[trim={0cm 0cm 0cm 0cm},clip,width=.9\textwidth]{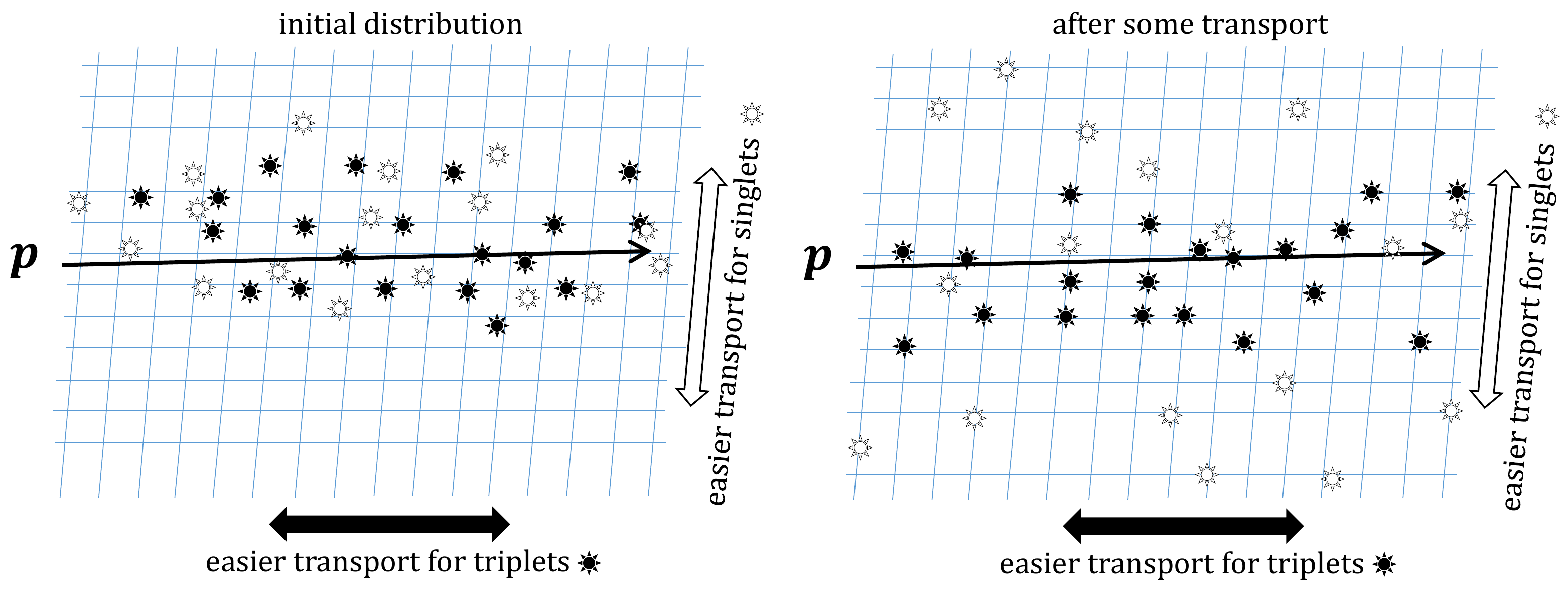}
	\caption{Illustration of excitation transport over time when excitations are populated by a proton recoil $p$ along a crystal axes with easy triplet (filled) transport and difficult singlet (open) transport. After some transport has occurred, the triplet excitation density remains high while the singlet density has dropped, leading to higher rates of TTA and lower rates of SIQ. This may lead to \Pave and \Dave being in sync. 
		\label{fig:anis_cartoon_2}}
\end{figure*}

\fig{fig:anis_cartoon} illustrates a scenario in which all excitations have the same directions of preferred transport. In this case, higher densities of singlet and triplet populations will occur in the same directions, so \Pave and \Dave are out of sync as higher rates of SIQ and TTA occur together. It is possible, however, that singlet and triplet excitations could have different directions of preferred transport. \fig{fig:anis_cartoon_2} illustrates such a scenario in which the excitations are populated along a direction with easier transport for triplets but more difficult transport for singlets. After some time passes, the triplets move within their initial distribution while the singlets move away from their initial distribution. This would produce higher rates of TTA and lower rates of SIQ than interactions at other directions, leading to more delayed light and more prompt light, so that \Pave and \Dave are in sync. 

In the two-dimensional space shown in Figs.~\ref{fig:anis_cartoon} and~\ref{fig:anis_cartoon_2}, the directions of maximum and minimum prompt and delayed light were paired either as in sync, where \Pmax and \Dmax are in the same direction and \Pmin and \Dmin are in the same direction, or as out of sync, where \Pmax and \Dmin are in the same direction and \Pmin and \Dmax are in the same direction. In p-terphenyl, however, \Pmin and \Dmin are in the same direction, but \Pmax and \Dmax are in different directions. This is possible because there are three crystal axes directions, and singlets and triplets will each have directions of easiest, moderate, and most difficult transport, so the maximal direction of one may correspond to the moderate direction of another. In fact, these three directions need not line up perfectly between singlets and triplets, as observed in anthracene, whose saddle point feature in \Pave is about \degrees{30} off from that the maximal feature in \Dave. 

In addition to preferred directions of excitation transport, there may be preferred directions for SIQ and TTA. Some previous authors have attributed the scintillation anisotropy entirely to preferred directions of transport~\cite{Heckmann1959,Brooks1974}, but we can not rule out preferred directions of SIQ and TTA interactions, as suggested in~\cite{Tsukada1962}.

The interactive kinetic processes are only directionally dependent on a bulk level in a crystalline material where the molecules follow orderly arrangement, which explains why no anisotropy is observed in amorphous plastic and liquid materials~\cite{Schuster2016a}. No anisotropy was observed for electron recoils produced by gamma-ray interactions or for muon interactions, both of which interact with low \dEdx~\cite{Schuster2016}. This indicates that the changes in excitation densities are only significant if the overall excitation density is already high, as produced by high \dEdx interactions by heavy charged particles such as a proton recoil or alpha particle. 
	
\subsection{Relationship to crystal structure}

The hypothesis presented in \secref{sec:hyp_anis} states that the anisotropy is produced by preferred directions of kinetic process(es), which may include excitation transport, SIQ, or TTA, within the crystal axes. These directional dependencies in the kinetic processes are due in large part to the molecular and crystal structure of each organic crystal, namely the spacing and relative orientation between molecules along each direction. In order to pursue understanding of what physical properties dictate the preferred directions, a few properties will be considered that are correlated with the relationship between \Pave and \Dave.

\begin{figure*}
	\centering
	\includegraphics[trim={0cm 0cm 0cm 0cm},clip,width=.85\textwidth]{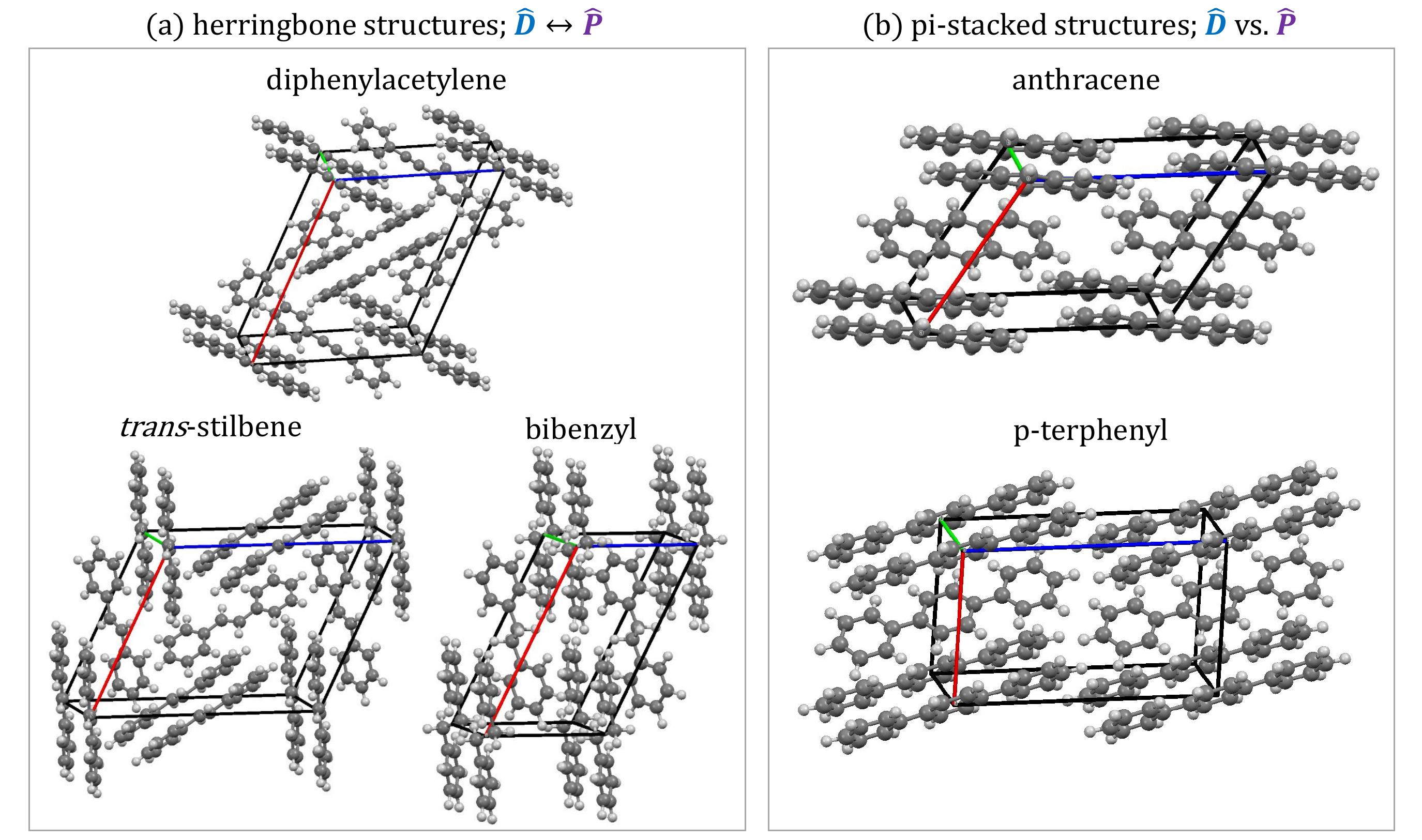}
	\caption{Crystal structures as visualized by Mercury software~\cite{Macrae2008}. (a) Diphenylacetylene, stilbene, and bibenzyl can be described as herringbone crystal structure and exhibit prompt and delayed light anisotropies that are in sync. (b) Anthracene and p-terphenyl both have $\pi$-stacked crystal structure and prompt and delayed light anisotropies that are out of sync. 
		\label{fig:xtal_structures}
	}
\end{figure*}

The first property to consider is the crystal structure of these materials. \fig{fig:xtal_structures} shows the crystal structures for all five materials measured in this paper. \fig{fig:xtal_structures}~(a) shows the three materials that presented an in sync relationship between the prompt and delayed light anisotropies: diphenylacetylene, stilbene, and bibenzyl. These materials all can be described as having herringbone crystal structure. \fig{fig:xtal_structures}~(b) shows the two materials that presented an out of sync relationship between the prompt and delayed light anisotropies: anthracene and p-terphenyl. These two materials both have a $\pi$-stacked crystal structure. 

This is the first time that the behavior of the effect has been correlated to the crystal structure. This observation strongly supports the theory that the crystal structure dictates the behavior of the effect. The hypothesis presented in \secref{sec:hyp_anis} is consistent with the hypothesis that the preferred directions of excitation transport for singlets and triplets are approximately the same in the $\pi$-stacked structures, making \Pave and \Dave out of sync, and different in the herringbone structures, making \Pave and \Dave in sync. This introduces the question of what structural property dictates the directions of preferred singlet and triplet transport and interactions. Considering excitation transport alone, the preferred direction of singlet transport depends on a dipole-dipole interaction, while the preferred direction of triplet transport depends on $\pi$-orbital overlap~\cite{Akselrod2014}. Both of these depend on the crystal and molecular structure, and may or may not be highest along the same crystal axes. Previous authors have also indicated that the electrical conductivity may play an important role~\cite{W.F.Kienzle1961}.

Considering the properties shown in \tab{tab:materials}, three other collective properties emerge that are correlated with the  behavior. First, the two materials with out of sync \Pave and \Dave, anthracene and p-terphenyl, were larger in size than the other materials. The size of each crystal is not expected to be the reason why the effect behaves differently in these materials, as previous work showed that the scintillation anisotropy in stilbene was approximately uniform across several samples of different sizes and shapes~\cite{Schuster2016a}. Second, these two materials were grown with melt-growth techniques, while the others were grown with solution processing. While it has been demonstrated that PSD performance can be significantly degraded in melt-grown crystals~\cite{Fraboni2016}, previous anisotropy measurements demonstrated that the anisotropy effect is roughly the same between melt-grown and solution-grown stilbene~\cite{Schuster2016a}. Thus, the effect of imperfections in the crystal structure that result in melt-grown crystals on the anisotropy effect does not appear to be strong. Third, these two materials have higher densities than the other three. While this is also not expected to be responsible for the anisotropy behavior, this may be a correlated physical property that results from the tighter $\pi$-stacked crystal structures.

\section{Summary}

This paper contributes new measurements and information on the scintillation anisotropy in organic crystal scintillators, making progress toward understanding the physical mechanisms that govern the effect. 

This paper presented experimental characterizations of the scintillation anisotropy in five pure organic crystal scintillators. A full hemisphere of angles were characterized at \MeV{14.1} and \MeV{2.5} in anthracene and stilbene and at \MeV{14.1} in p-terphenyl, bibenzyl, and diphenylacetylene. These measurements demonstrate that the magnitude and behavior of the scintillation anisotropy varies across materials and is correlated with crystal structure. The effect is consistent with directionally-dependent kinetic processes for the $\pi$-molecular excitations including singlet ionization quenching, triplet-triplet annihilation, and excitation transport. 

Opportunities exist to continue this investigation, starting with continued measurements of radiation interactions. One could measure the scintillation anisotropy in additional pure crystalline materials and see whether the correlations to crystal structure are consistent with those presented in this paper. Also, one could measure the effect at additional energies to investigate the interesting relationship to proton recoil energy. The dependence on molecular and crystal structure is likely not straight forward, but experimental or quantitative methods may exist for calculating the preferred directions of singlet and triplet transport. It may be possible to determine the direction of preferred singlet transport, a longer range interaction, by measuring the photoconductivity across a bulk material in different directions. Triplet transport, a shorter range interaction, depends more on the local crystal and electronic structure of a molecule. Optical measurement techniques exist for measuring the directional diffusion length of triplet excitations, such as that used by Akselrod~\etal on tetracene~\cite{Akselrod2014}. Also, computational methods such as density functional theory (DFT) may be useful in calculating the directional transport likelihood for triplet excitations.

\section*{Acknowledgements}

The authors thank Natalia Zaitseva at Lawrence Livermore National Laboratory for providing several samples for characterization and John Steele for his assistance in building the apparatus. 

\small This work is supported in part by the University of Michigan. This material is based upon work supported by the National Science Foundation Graduate Research Fellowship Program under Grant no. DGE 1106400. This material is based upon work supported by the Department of Energy National Nuclear Security Administration under Award Number DE-NA0000979 through the Nuclear Science and Security Consortium. Sandia National Laboratories is a multimission laboratory managed and operated by National Technology and Engineering Solutions of Sandia LLC, a wholly owned subsidiary of Honeywell International Inc. for the U.S. Department of Energy’s National Nuclear Security Administration under contract DE-NA0003525.

\printbibliography

\end{document}